\documentclass[twocolumn,english,amsart,showpacs,preprintnumbers,amsmath,amssymb,floatfix]{revtex4-1}
\usepackage{tikz,xcolor}
\usepackage{amsmath}
\usepackage[colorlinks = true,
linkcolor = blue,
urlcolor  = blue,
citecolor = blue,
anchorcolor = blue]{hyperref}

\definecolor{lime}{HTML}{A6CE39}
\DeclareRobustCommand{\orcidicon}{%
	\begin{tikzpicture}
		\draw[lime, fill=lime] (0,0)
		circle [radius=0.16]
		node[white] {{\fontfamily{qag}\selectfont \tiny ID}};
		\draw[white, fill=white] (-0.0625,0.095)
		circle [radius=0.007];
	\end{tikzpicture}
	\hspace{-2mm}
}

\foreach \x in {A, ..., Z}{%
	\expandafter\xdef\csname orcid\x\endcsname{\noexpand\href{https://orcid.org/\csname orcidauthor\x\endcsname}{\noexpand\orcidicon}}
}

\usepackage[T1]{fontenc}
\usepackage[latin9]{inputenc}
\usepackage{color}
\usepackage{array}
\usepackage{amstext}
\usepackage{graphicx}
\usepackage{esint}
\usepackage{rotating}
\usepackage{appendix}
\usepackage{float}
\usepackage{booktabs}

\usepackage[font=small,labelfont=bf]{caption}
\usepackage{xcolor}
\usepackage{ulem}

\makeatletter

\@ifundefined{textcolor}{}
{%
	\definecolor{BLACK}{gray}{0}
	\definecolor{WHITE}{gray}{1}
	\definecolor{RED}{rgb}{1,0,0}
	\definecolor{GREEN}{rgb}{0,1,0}
	\definecolor{BLUE}{rgb}{0,0,1}
	\definecolor{CYAN}{cmyk}{1,0,0,0}
	\definecolor{MAGENTA}{cmyk}{0,1,0,0}
	\definecolor{YELLOW}{cmyk}{0,0,1,0}
}

\@ifundefined{definecolor}
{\usepackage{color}}{}
\@ifundefined{definecolor} 
{\usepackage{color}}{}
\makeatother
\usepackage{babel}

\allowdisplaybreaks

\begin{document}
	
	%%%%%%%%%%%%%%%%%%%%%%%%%%%%%%

\title{From QCD-Based Descriptions to Direct Fits: A Unified Study of Nucleon Electromagnetic Form Factors}

	\author{Hossein Vaziri$^{1}$\orcidA{}}
\email{Hossein.Vaziri@shahroodut.ac.ir}

\author {Mohammad Reza Shojaei$^{2}$\orcidB{}}
\email{Shojaei.ph@gmail.com}

\author{Pere Masjuan$^{3}$\orcidC{}}
\email{masjuan@ifae.es}

\affiliation {
	$^{(1,2)}$Department of Physics, Shahrood University of Technology, P. O. Box 36155-316, Shahrood, Iran }

\affiliation{
	$^{(3.a)}$Departamento de F\'{\i}sica d'Altes Energies (IFAE) and
	The Barcelona Institute of Science and Technology (BIST),
	Campus UAB, 08193 Bellaterra (Barcelona), Spain}

\affiliation{
	$^{(3.b)}$Grup de F\'isica Te\`orica, Departament de F\'isica,
	Universitat Aut\`onoma de Barcelona, The Barcelona Institute of Science and Technology (BIST),and Institut de F\'isica d'Altes Energies (IFAE),08193 Bellaterra (Barcelona), Spain}

	\date{\today}

	%
	%%%%%%%%%%%%%%%%%%%%%%%%%%%%%%%%%%%%%%%%%%%%%  Abstract    %%%%%%%%%%%%%%%%%%%%%%%%%%%%%%%%%%%%%%%%%%%%%%%%%%%%%%%%%%
	%

\begin{abstract}\label{abstract}
	We present a detailed study of the nucleon electromagnetic form factors in the spacelike region by combining three complementary approaches: two GPD-based contributions and a vector-meson exchange component. By fitting experimental data, we extract the optimal weights and shape parameters describing the proton and neutron form factors.  
	Global Pad\'e-based fits are then constructed for four distinct groups of form factors, starting from local Taylor expansions and yielding stable analytic parametrizations over the analyzed $t$ range. The combined framework provides an accurate and physically motivated description of nucleon structure within a controlled model-dependent setting across a wide range of momentum transfers.
\end{abstract}

	%
	
	%\pacs{12.38.-t, 24.85.+p, 13.15.+g, 13.60.Hb}
	
	\maketitle

    %%%%%%%%%%%%%%%%%%%%%%%%%%%%%%%%%%%%%%%%%%%%%%%%%
\section{INTRODUCTION}\label{sec:sec1} 
Hadrons are composite systems made of strongly interacting quarks and gluons, described by the theory of Quantum Chromodynamics (QCD)~\cite{Drechsel:2007sq,Miller:1990iz,JeffersonLabHallA:2023rsh,Radyushkin:2011dh,Drell:1969km}. Understanding the internal structure of nucleons remains a central goal in hadronic physics. Electromagnetic form factors serve as essential tools in this regard, encoding information about the spatial and spin distributions of charge and magnetization inside nucleons. These form factors are accessible through electron-nucleon scattering experiments, notably using the Rosenbluth separation method~\cite{Halzen, Rosenbluth:1950yq}. They are defined as matrix elements of the electromagnetic current between nucleon states with different momenta and exhibit a characteristic falloff with increasing momentum transfer $Q^2$~\cite{Gutbrod:1978sb,Martin:2009iq,RezaShojaei:2016oox,Manohar:1998xv,SattaryNikkhoo:2018odd}.

Two primary theoretical approaches are commonly employed for computing form factors. One is based on Generalized Parton Distributions (GPDs), which provide a unified description of parton distributions and form factors by incorporating the variables $x$ (Bjorken scaling), $t$ (momentum transfer), and $\xi$ (skewness)~\cite{Selyugin:2023hqu, SattaryNikkhoo:2018gzm, HajiHosseiniMojeni:2022okc,Nikkhoo:2015jzi,HajiHosseiniMojeni:2022tzn,SattaryNikkhoo:2018gzm,Diehl:2003ny,Radyushkin:1997ki,Guidal:2013rya}. Another approach is the vector meson-dominance (VMD) model, where vector mesons mediate the interaction between the photon and nucleon~\cite{Sakurai:1960ju}. This method effectively describes the low-$Q^2$ behavior of form factors, but since it depends only on $t$, it has limited applicability at higher $Q^2$ where partonic degrees of freedom become significant.
In this work, we propose four combined models composed of three components: (i) a VMD model~\cite{Masjuan:2012sk}, (ii) a GPD model in combination with the VS24 ansatz~\cite{Vaziri:2024fud,Vaziri:2023xee,vaziri:2025xkw} and the KKA10 parton distribution functions (PDFs)~\cite{Khorramian:2009xz}, and (iii) a GPD model based on the ER ansatz~\cite{Guidal:2004nd} in combination with the MRST2002 PDFs~\cite{ Martin:2002dr}. Each model contributes to the final form factors through a fitted weight parameter, allowing us to quantify the effective role of each model component across different $Q^2$ regions. These three frameworks were selected because they complement each other: the VMD model captures low-energy vector meson exchange physics, while the GPD-based components describe the partonic structure and its evolution with $t$. By fitting the combined model to experimental data~\cite{Qattan:2012zf}, we determine the best parameter values and analyze how the Dirac and Pauli form factors behave for both proton and neutron. This method improves the fit quality and helps us understand the contribution of different models in various $Q^2$ regions, especially the balance between low and high-energy effects in nucleon structure. In addition to this combined framework, we further perform dedicated Pad\'e approximant-based global parametrizations to provide stable analytic representations of the form factors across different kinematic groups.

The paper is organized as follows. In Sec.~\ref{sec:sec2}, we describe the VMD approach in the large-$N_c$ limit of QCD. Sec.~\ref{sec:sec3} covers the GPD formalism and the ansatz used. The construction and fitting of the combined models are presented in Sec.~\ref{sec:sec4}. Phenomenological fits for neutron form factors using Regge-inspired and Pad\'e approximants forms are discussed in Sec.~\ref{sec:sec5}. Global Pad\'e approximants functions and numerical analysis for four groups of form factors are obtained in Sec.~\ref{sec:sec6}, where the results are also compared with experimental data. The conclusions are summarized in Sec.~\ref{sec:conclusion}.

\newpage

\section{Meson Dominance and Large-\textit{N}\textsubscript{c} Approach to Electromagnetic Nucleon Form Factors}\label{sec:sec2}
     Vector mesons play a central role in the Vector Meson Dominance (VMD) model, which provides a physical interpretation of how electromagnetic interactions probe the internal structure of hadrons. In the VMD picture, the photon interacts with the nucleon not directly, but through conversion into a vector meson. This mechanism helps describe the behavior of nucleon electromagnetic form factors over different momentum transfers~\cite{Sakurai:1960ju}.

     In the large-$N_c$ limit of QCD, where mesons become narrow and stable, the nucleon form factors can be expressed as sums over vector-meson poles. In the isospin-symmetric limit with light $u,d$ (and a spectator $s$), the electromagnetic current reads
    
     \begin{equation}
     	J^\mu_{\rm em}(x)=\bar q(x)\,Q\,\gamma^\mu q(x)
     	= \underbrace{\frac{1}{6}\,\bar q(x)\gamma^\mu q(x)}_{J^{(0)\mu}(x)}
     	+ \underbrace{\bar q(x)\gamma^\mu\frac{\tau^3}{2}q(x)}_{J^{(1)\mu}(x)},
     	\label{eq:jem}    
\end{equation}
     i.e. an isoscalar piece $J^{(0)\mu}$ and an isovector piece $J^{(1)\mu}$. Their matrix elements define the isoscalar/isovector Dirac and Pauli form factors:

     \begin{equation}
     	\begin{aligned}
     		\langle N(p')| J^{(0)\mu}(0) |N(p)\rangle
     		&= \bar u(p') \Big[
     		\gamma^\mu F_1^{(0)}(t) \\
     		&\quad + i \frac{\sigma^{\mu\nu} q_\nu}{2 M_N} F_2^{(0)}(t)
     		\Big] u(p),\\[4pt]
     		\langle N(p')| J^{(1)\mu}(0) |N(p)\rangle
     		&= \bar u(p') \Big[
     		\gamma^\mu F_1^{(1)}(t) \\
     		&\quad + i \frac{\sigma^{\mu\nu} q_\nu}{2 M_N} F_2^{(1)}(t)
     		\Big] \frac{\tau^3}{2} u(p),
     	\end{aligned}
     	\label{eq:Matrix}
     \end{equation}

     So that physical proton and neutron form factors follow from
     \begin{equation}
     	F_{1,2}^{p}(t)=F_{1,2}^{(0)}(t)+F_{1,2}^{(1)}(t),\qquad
     	F_{1,2}^{n}(t)=F_{1,2}^{(0)}(t)-F_{1,2}^{(1)}(t).
     	\label{eq:ffpn}
     \end{equation}
     
     At $t=0$ the normalizations reproduce the electric charges and anomalous magnetic moments of proton and neutron as:
     \begin{equation}
     	F_1^p(0)=1,\quad F_1^n(0)=0,\qquad
     	F_2^p(0)=\kappa_p,\quad F_2^n(0)=\kappa_n,
     	\label{eq:f1f2norm}
     \end{equation}
     with $\kappa_p=1.793$ and $\kappa_n=-1.913$.
     
     Throughout the paper we adopt the spacelike convention $t\equiv Q^2>0$. In terms of Dirac/Pauli form factors, the Sachs form factors are then \cite{Masjuan:2012sk}
    
     \begin{equation}
     	G_E^{p,n}(t)=F_1^{p,n}(t)-\frac{t}{4M_N^2}\,F_2^{p,n}(t),
     	\label{eq:sachs1}
     \end{equation}
 
          \begin{equation}
            	G_M^{p,n}(t)=F_1^{p,n}(t)+F_2^{p,n}(t).
            	     	\label{eq:sachs2}
        \end{equation}
     
     Convenient large-$N_c$ representation uses products of monopoles \cite{Masjuan:2012sk},
     \begin{equation}
     	\begin{aligned}
     		F_1^{(0)}(t) &= \frac{1 - c_0\, t/m_{\omega''}^2}
     		{(1 - t/m_\omega^2)(1 - t/m_{\omega'}^2)(1 - t/m_{\omega''}^2)},\\
     		F_1^{(1)}(t) &= \frac{1 - c_1\, t/m_{\rho''}^2}
     		{(1 - t/m_\rho^2)(1 - t/m_{\rho'}^2)(1 - t/m_{\rho''}^2)},\\
     		F_2^{(0)}(t) &= \frac{1}
     		{(1 - t/m_\omega^2)(1 - t/m_{\omega'}^2)(1 - t/m_{\omega''}^2)},\\
     		F_2^{(1)}(t) &= \frac{1}
     		{(1 - t/m_\rho^2)(1 - t/m_{\rho'}^2)(1 - t/m_{\rho''}^2)},
     	\end{aligned}
     	\label{eq:Fmodel}
     \end{equation}
     with $c_0$ and $c_1$ fixed by vector-meson nucleon couplings \cite{Masjuan:2014sua},
     \begin{equation}
     	\begin{aligned}
     		\frac{g_{\omega NN} f_{\omega\gamma}}{m_\omega^2}
     		&= \frac{1}{2}\,\frac{1 - c_0\, m_\omega^2/m_{\omega''}^2}
     		{(1 - m_\omega^2/m_{\omega'}^2)(1 - m_\omega^2/m_{\omega''}^2)},\\
     		\frac{g_{\rho NN} f_{\rho\gamma}}{m_\rho^2}
     		&= \frac{1}{2}\,\frac{1 - c_1\, m_\rho^2/m_{\rho''}^2}
     		{(1 - m_\rho^2/m_{\rho'}^2)(1 - m_\rho^2/m_{\rho''}^2)}.
     	\end{aligned}
     	\label{eq:c0c1}
     \end{equation}

 \section{GENERALIZED PARTON DISTRIBUTIONS
}\label{sec:sec3}

GPDs unify and extend the concepts of form factors and parton distributions, encoding both the spatial and momentum structure of hadrons. GPDs appear naturally in processes like deeply virtual Compton scattering (DVCS), where a virtual photon interacts with a quark inside the nucleon~\cite{Guidal:2004nd,Burkardt:2002hr}.

The hadron form factors are linked to the Generalized Parton Distributions (GPDs), which are functions of $(x, \xi, t)$, through the sum rules~\cite{Guidal:2004nd, Selyugin:2009ic}:

  \begin{equation}
  	F_{1}(t)=\sum_{q} e_{q}\int_{-1}^{1}dx {H}^{q}(x,t,\xi),
  	\label{eq:f0}
  \end{equation}

  \begin{equation}
  	F_{2}(t)=\sum_{q} e_{q}\int_{-1}^{1}dx {E }^{q}(x,t,\xi),
  	\label{eq:f2}
  \end{equation}
In the spacelike region, where the momentum transfer is purely transverse, we set $\xi = 0$~\cite{Guidal:2004nd}.
 Researchers have revised elastic form factors to reduce the integration region within the $0 < x < 1$ range, resulting in alternative expressions~\cite{Martin:2002dr}:

 \begin{equation}
	F_{1}(t)=\sum_{q} e_{q}\int_{0}^{1}dx \, \mathcal{H}^{q}(x,t,\xi=0),
	\label{eq:f1}
\end{equation}

\begin{equation}
	F_{2}(t)=\sum_{q} e_{q}\int_{0}^{1}dx \, \varepsilon^{q}(x,t,\xi=0),
	\label{eq:f22}
\end{equation}

At $t \to 0$, the functions $\mathcal{H}^q(x,t)$ reduce to the usual valence quark densities~\cite{Guidal:2004nd},
\\
  \begin{equation}
\mathcal{H}^u(x,t=0)= u_v(x) ,  \,\,\,\,\,\,\,\,\,\,\    \mathcal{H}^d(x,t=0)= d_v(x),
	\label{eq:1}
  \end{equation}

with the integrals\\

  \begin{equation}
\int_{0}^{1}  u_v(x) dx=2 ,\,\,\,\,\,\,\,\,\,\,\    \int _{0}^{1} d_v(x) dx=1 .
	\label{eq:2}
  \end{equation}
normalized to the proton's $u$ and $d$ valence quark numbers. These conditions ensure consistency with the form factor normalizations at $t=0$ already stated in Sec.~\ref{sec:sec2}.

The functional forms of $\mathcal{H}(x)$ and $\varepsilon(x)$ vary among different models. To achieve a faster falloff with $t$, $\varepsilon(x)$ must include additional powers of $(1-x)$ at large $x$, compared to $\mathcal{H}(x)$~\cite{Guidal:2004nd,Selyugin:2009ic,Mojeni:2020rev,Shojaei:2015oia,vaziri:2025xkw}. Hence, we have:
\begin{eqnarray}
	\varepsilon _{u}(x) &=&\frac{\kappa _{u}}{N_{u}}(1-x)^{\eta _{u}}u_{v}(x), \nonumber\\
	\varepsilon _{d}(x) &=&\frac{\kappa _{d}}{N_{d}} (1-x)^{\eta _{d}}d_{v}(x),\label{eq:Eud1}
\end{eqnarray}

\begin{equation}
	\kappa _{q}=\int_{0}^{1}dx\varepsilon _{q}(x).
\end{equation}

The Pauli form factors must satisfy $F_2^p(0)=\kappa_p = 1.793$ and $F_2^n(0)=\kappa_n = -1.913$, which imposes constraints on $\kappa_{q}$:

\begin{eqnarray}
	\kappa_u=\kappa_n+2\kappa_p,\nonumber\\
	\kappa_d=2\kappa_n+\kappa_p.
	\label{eq:kapp}
\end{eqnarray}

where the normalization factors $N_u$ and $N_d$ are determined as~\cite{Guidal:2004nd}:

\begin{eqnarray}
	N_{u} &=&\int_{0}^{1}dx(1-x)^{\eta _{u}}u_{v}(x), \\
	N_{d} &=&\int_{0}^{1}dx(1-x)^{\eta _{d}}d_{v}(x).  \nonumber
	\label{eq:Nud}
\end{eqnarray}

In order to satisfy Eq.~(\ref{eq:Eud1}), the nucleon form factor data can be fitted to obtain the values of $\eta_u$ and $\eta_d$.  

The extended ER ansatz~\cite{Guidal:2004nd}, the MRST2002 approximation~\cite{Martin:2002dr}, and the VS24 ansatz~\cite{Vaziri:2024fud,vaziri:2025xkw} with different $t$-dependencies will be used in our analysis. The {\bf ER} ansatz reads~\cite{Guidal:2004nd} 

\begin{equation}
	\mathcal{H}^{q}(x,t)=q_{v}(x)x^{-\alpha ^{\prime }(1-x)t},
	\label{eq:HER}
\end{equation}
\begin{equation}
	\varepsilon _{q}(x,t)=\varepsilon _{q}(x)x^{-\alpha ^{\prime }(1-x)t}.
	\label{eq:EER}
\end{equation}

\begin{figure*}
	%\vspace*{-.658cm}
	\includegraphics[clip,width=0.45\textwidth]{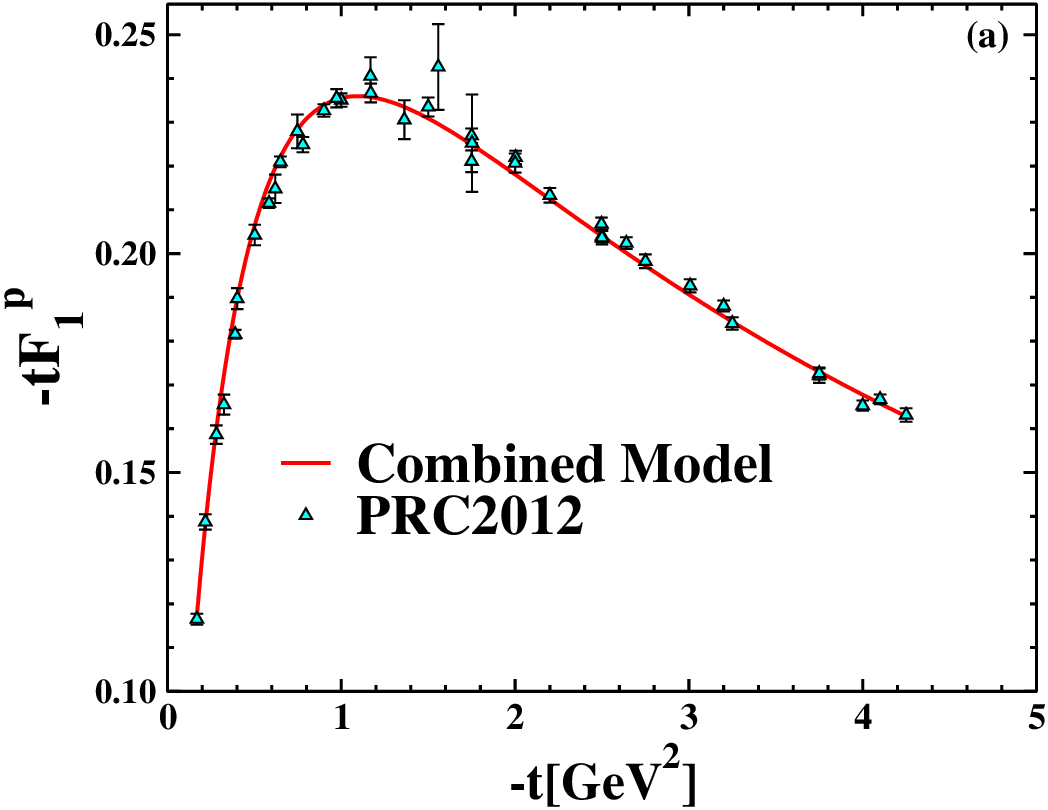}
	\hspace*{2mm}
	\includegraphics[clip,width=0.45\textwidth]{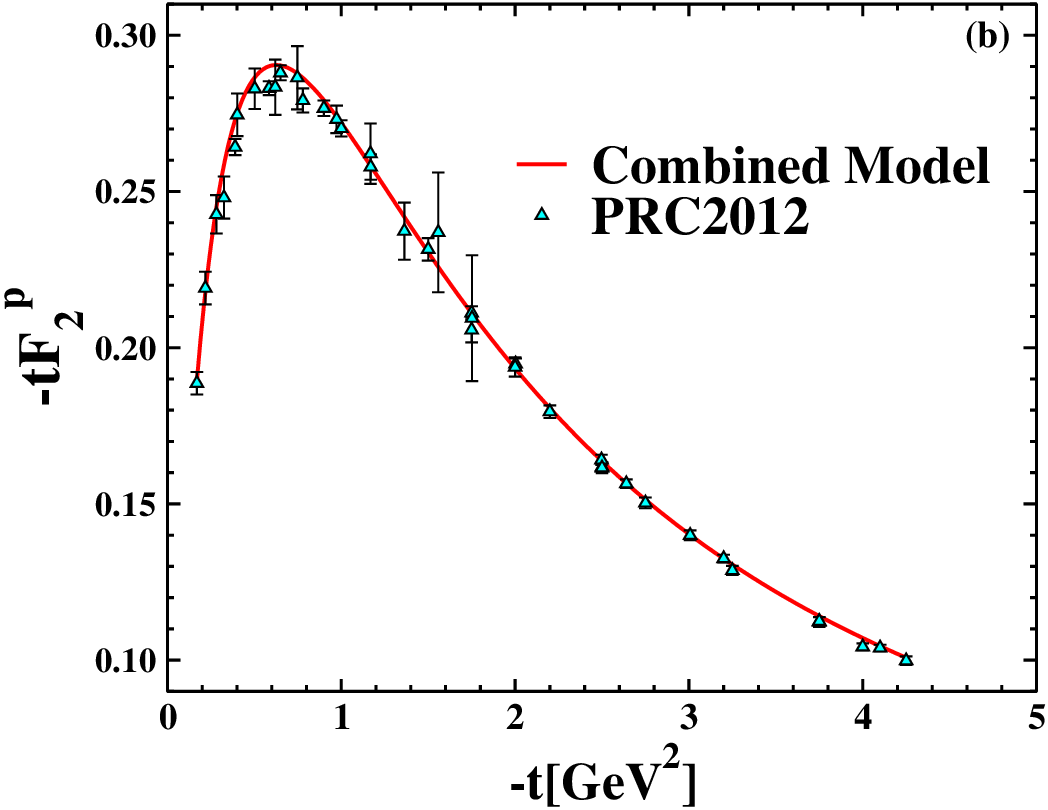}
	\vspace*{3mm}\\
	\includegraphics[clip,width=0.45\textwidth]{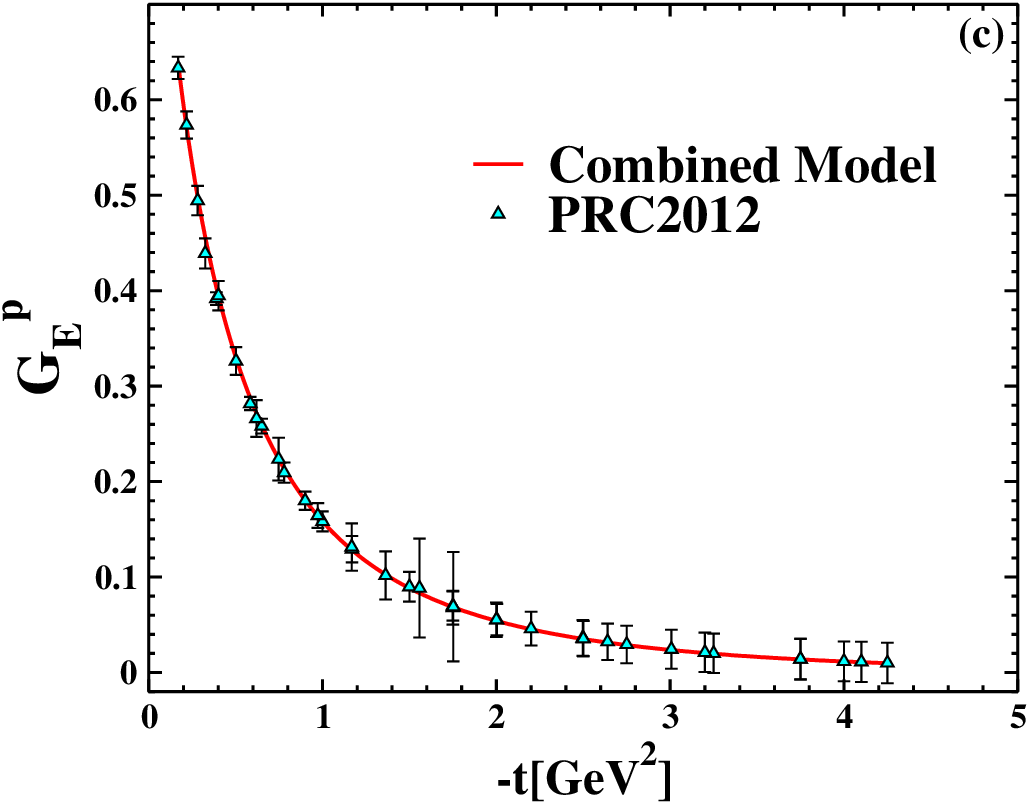}
	\hspace*{2mm}
	\includegraphics[clip,width=0.45\textwidth]{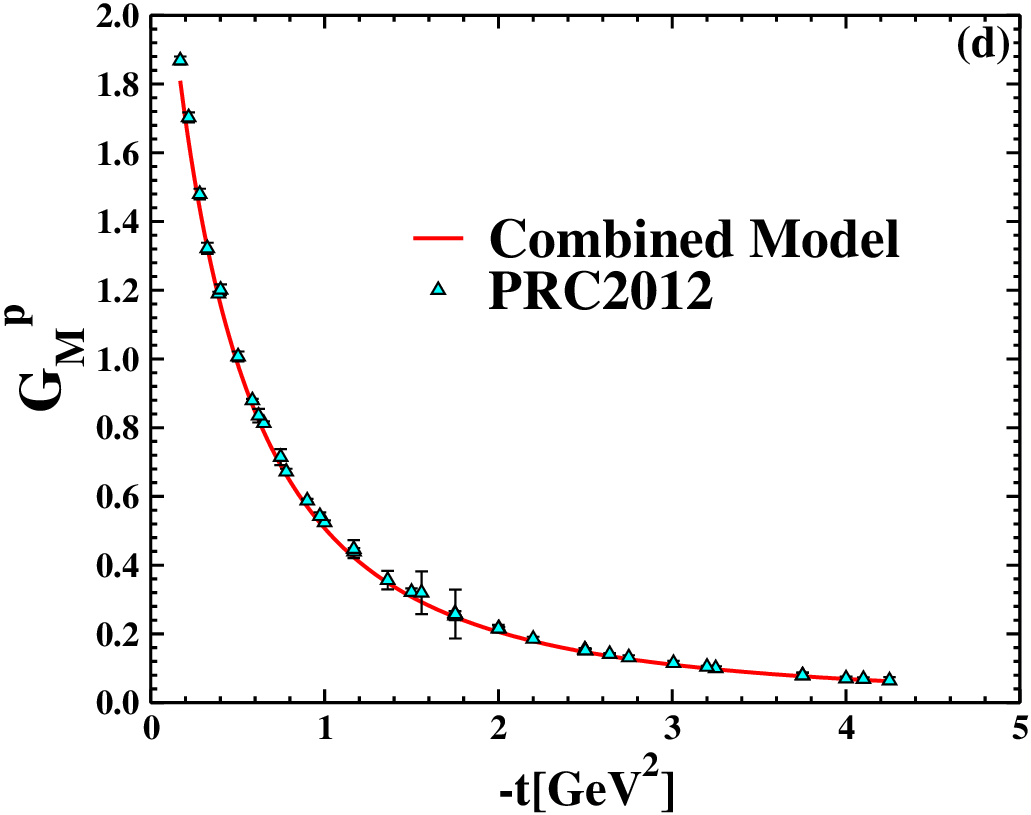}
	\vspace*{1.5mm}
	
	\caption{\footnotesize  The combined model of $tF_1^{p}$, $tF_2^{p}$, $G_E^{P}$, and $G_M^{p}$ (group 1) as a function of $-t$. The ER Ansatz~\cite{Guidal:2004nd}+MSRT2002 PDF~\cite{Martin:2002dr}, the VS24 Ansatz~\cite{Vaziri:2024fud} with new parameters+KKA10 PDF~\cite{Khorramian:2009xz}, and the VMD model~\cite{Masjuan:2012sk} are used. We fit Combined model to the nucleon form factors, whose numerical values were extracted from experimental data and reported by the authors in Ref.
		~\cite{Qattan:2012zf}.}
	\label{fig:Proton}
\end{figure*}

\begin{figure*}
	%\vspace*{-.658cm}
	\includegraphics[clip,width=0.45\textwidth]{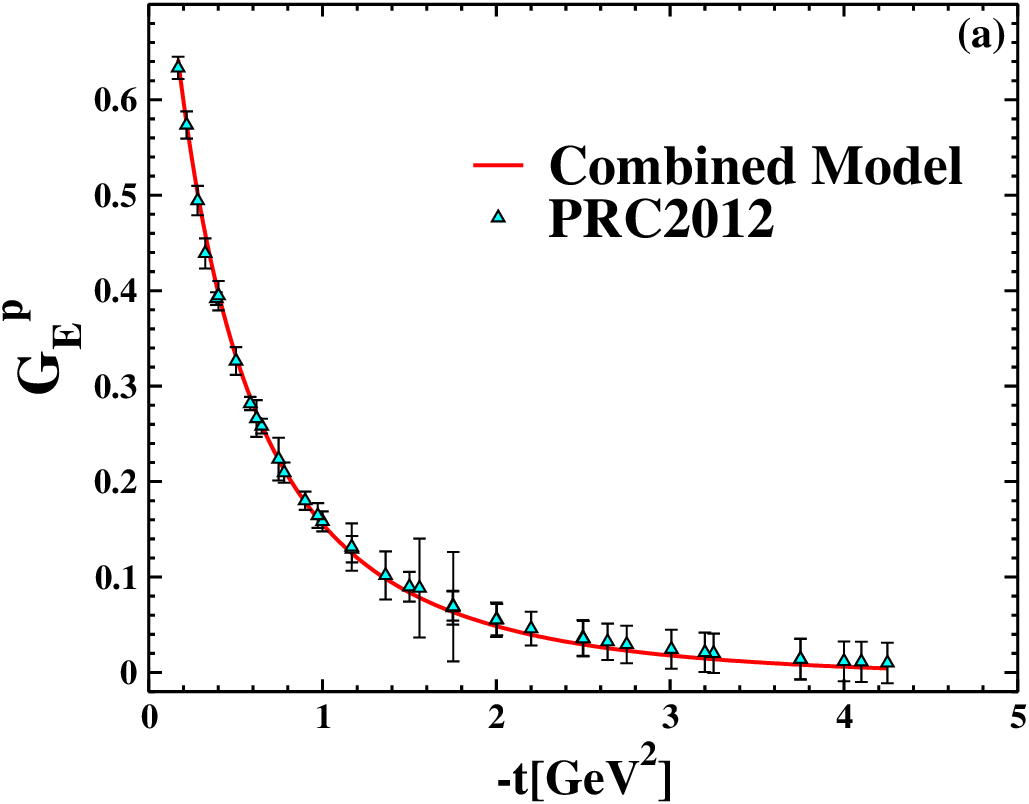}
	\hspace*{2mm}
	\includegraphics[clip,width=0.45\textwidth]{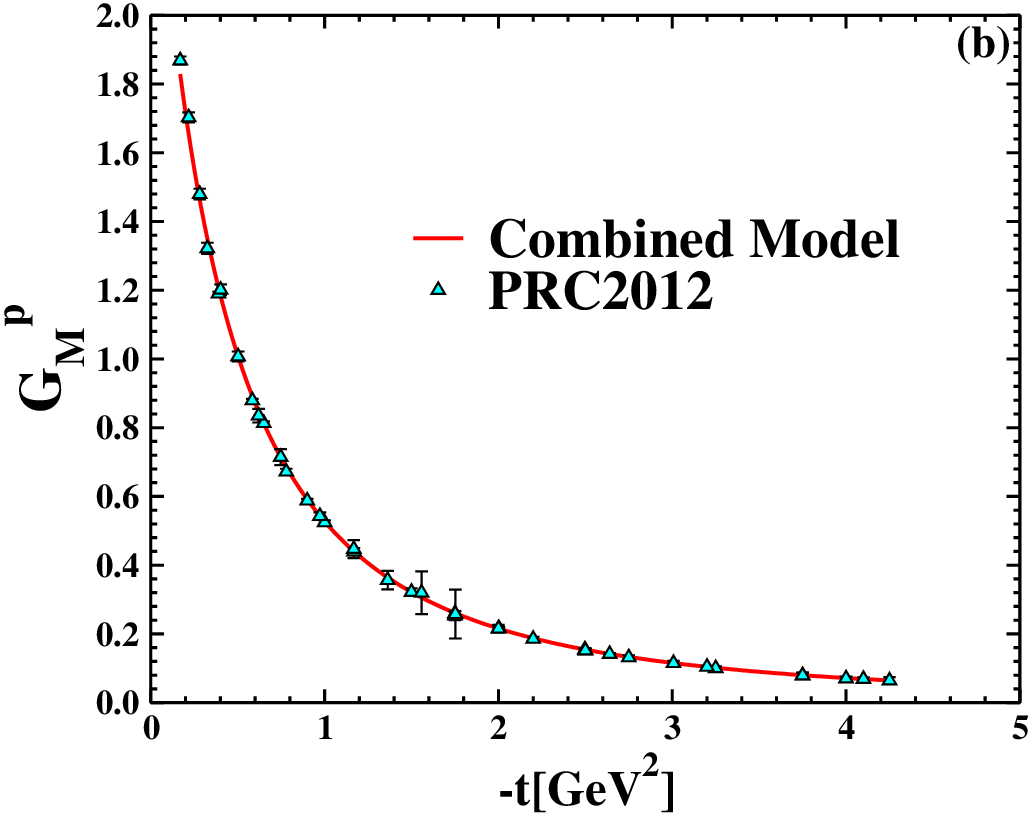}
	\vspace*{3mm}\\
	\includegraphics[clip,width=0.45\textwidth]{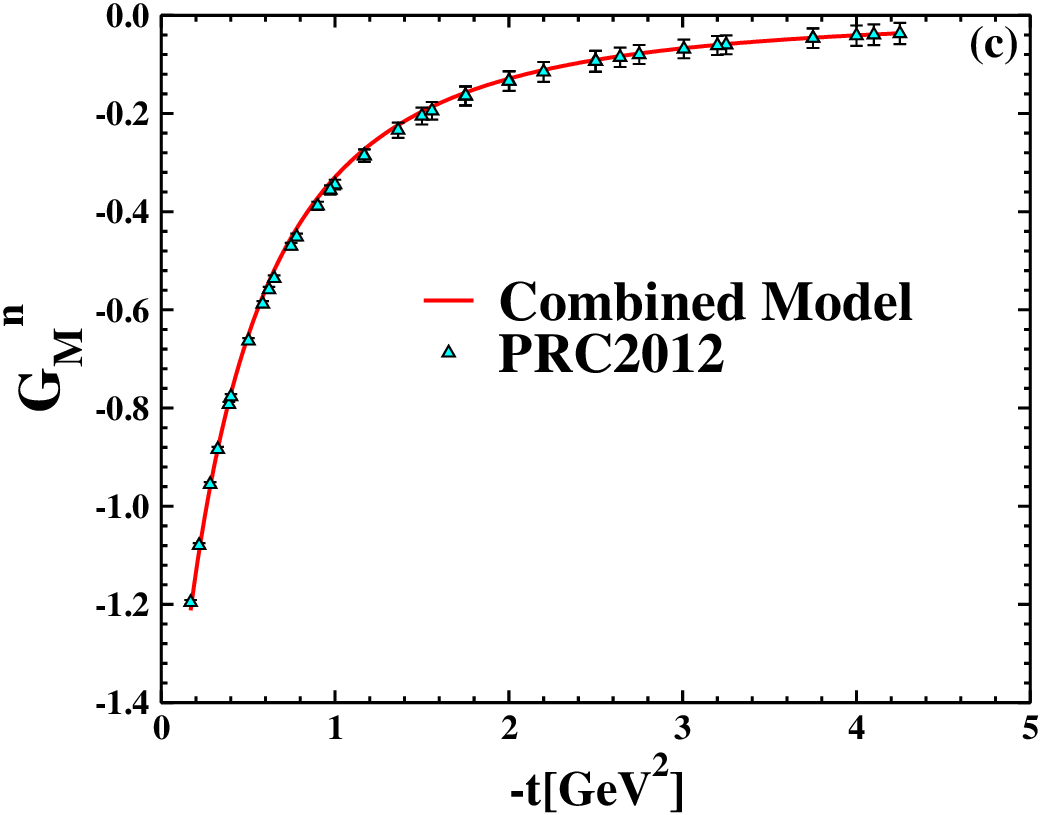}
	\vspace*{3mm}\\
	
	\	\caption{\footnotesize  The combined model of  $G_E^{p}$,  $G_M^{p}$ and $G_M^{n}$ (group 2) as a function of $-t$. The ER Ansatz~\cite{Guidal:2004nd}+MSRT2002 PDFs~\cite{Martin:2002dr}, the VS24 Ansatz~\cite{Vaziri:2024fud} with new parameters+KKA10 PDF~\cite{Khorramian:2009xz}, and the VMD model~\cite{Masjuan:2012sk} are used. We fit Combined model to the nucleon form factors, whose numerical values were extracted from experimental data and reported by the authors in Ref.
		~\cite{Qattan:2012zf} (upward triangles).
	}
	\label{fig:GGG}
\end{figure*}

\begin{figure*}
	%\vspace*{-.658cm}
	\includegraphics[clip,width=0.45\textwidth]{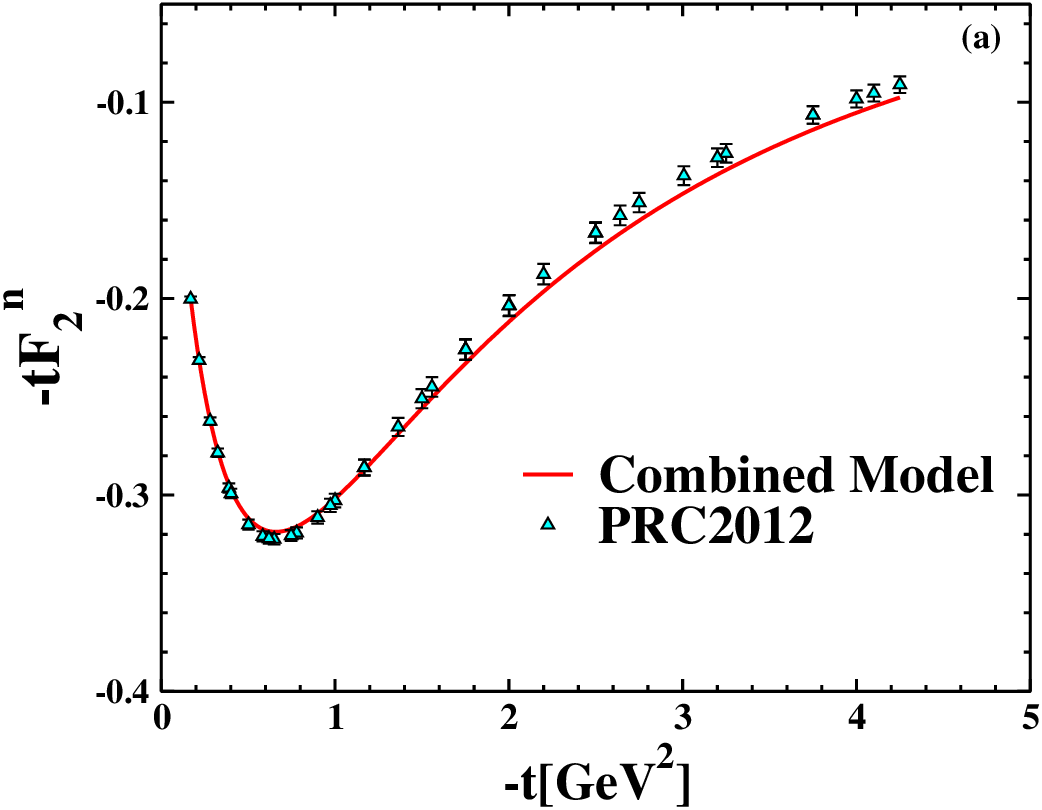}
	\hspace*{2mm}
	\includegraphics[clip,width=0.45\textwidth]{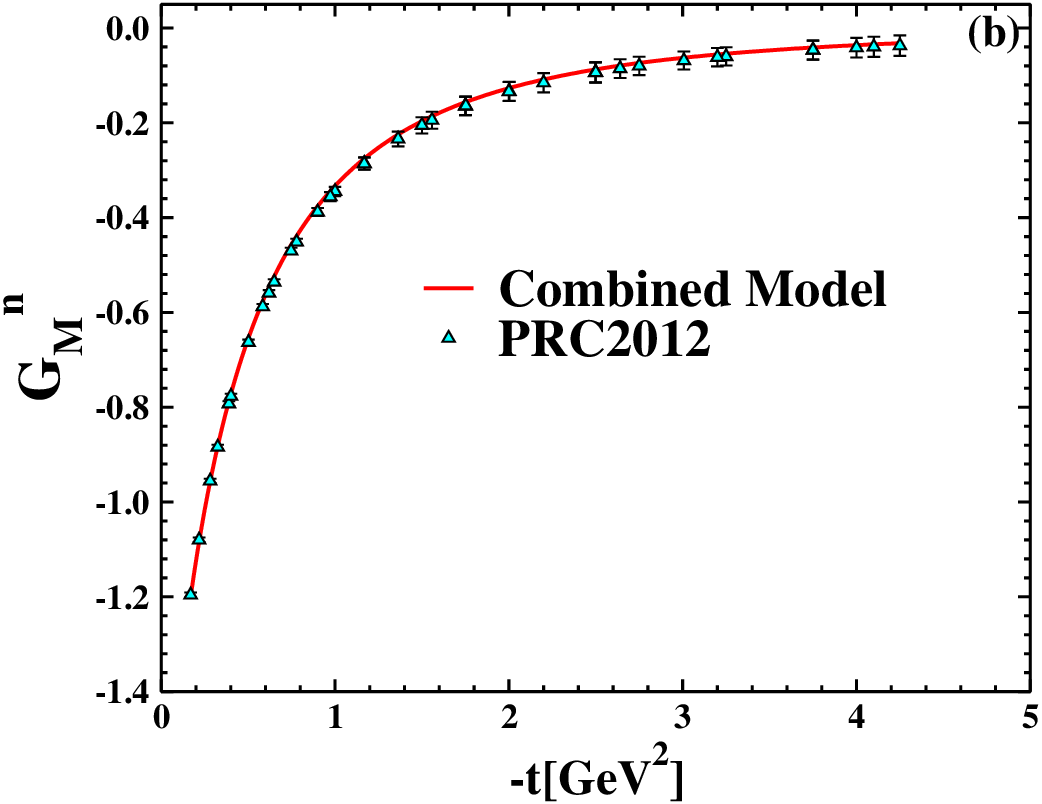}
	\vspace*{3mm}\\
	
	\	\caption{\footnotesize  The combined model of  $tF_2^{n}$ and $G_M^{n}$ (group 3) as a function of $-t$. The ER Ansatz~\cite{Guidal:2004nd}+MSRT2002 PDFs~\cite{Martin:2002dr}, the VS24 Ansatz~\cite{Vaziri:2024fud} with new parameters+KKA10 PDF~\cite{Khorramian:2009xz}, and the VMD model~\cite{Masjuan:2012sk} are used. We fit Combined model to the nucleon form factors, whose numerical values were extracted from experimental data and reported by the authors in Ref.
		~\cite{Qattan:2012zf}.}
	\label{fig:neutron1}
\end{figure*}

  The free parameter for the  extended ER ansatz is $\alpha^{\prime}=1.15$ from Ref. \cite{Masjuan:2012gc}.
The {\bf MRST2002} parton distribution functions (PDFs) at the $N^2LO$ approximation are given by~\cite{Martin:2002dr}:
\begin{equation}
	u_v(x)=0.262x^{-0.69}(1-x)^{3.5}(1+3.83x^{0.5}+37.65x),
	\label{eq:KUV1}
\end{equation}

\begin{equation}
	d_v(x)=1.061x^{-0.65}(1-x)^{4.03}(1+49.05x^{0.5}+8.65x).
	\label{eq:KdV1}
\end{equation}

We introduced the {\bf VS24} Ansatz~\cite{Vaziri:2024fud}, defined as:

\begin{equation}
	\mathcal{H}_{q}(x,t)=q_{v}(x)\;\exp [-\alpha^{\prime \prime } t(1-x)^\gamma\ln (x)+\beta x^{m^\prime}\ln (1-bt)],
	\label{eq:H}
\end{equation}
\begin{equation}
	\varepsilon _{q}(x,t)=\varepsilon _{q}(x)\;\exp [-\alpha^{\prime \prime } t (1-x)^\gamma\ln (x)+\beta
	x^{m^\prime}\ln (1-bt)],
	\label{eq:E}
\end{equation}
 The {\bf KKA10} parton distribution functions at $N^3LO$ approximation are~\cite {Khorramian:2009xz}:
\begin{equation}
	xu_v=3.41356 x^{0.298}(1-x)^{3.76847}(1+0.1399x^{0.5}-1.12x),
	\label{eq:6}
\end{equation}

\begin{equation}
	xd_v=5.10129x^{0.79167}(1-x)^{4.02637}(1+0.09x^{0.5}+1.11x).
	\label{eq:7}
\end{equation}
Using the VS24~\cite{Vaziri:2024fud} and ER ansatzes~\cite{Guidal:2004nd} in combination with  KKA10~\cite{Khorramian:2009xz} and MRST2002~\cite{Martin:2002dr} PDFs, respectively, we can calculate the form factors of the $u$ and $d$ quarks based on the formalism described earlier.  

The proton and neutron Dirac form factors are defined as~\cite{Ernst:1960zza,Blumlein:2006be,Blumlein:2021lmf}:
\begin{equation}
	F_{1}^{p}(t)= e_u F_{1}^{u}(t)+e_d F_{1}^{d}(t)  ,
		\label{eq:p1}
\end{equation}

\begin{equation} 
F_{1}^{n}(t)= e_u F_{1}^{d}(t)+e_d F_{1}^{u}(t).
\label{eq:n1}
\end{equation}

Accordingly, the Pauli form factors are given by:
\begin{equation}
	F_{2}^{p}(t)= e_u F_{2}^{u}(t)+e_d F_{2}^{d}(t)  ,
		\label{eq:p2}
\end{equation}

\begin{equation} 
	F_{2}^{n}(t)= e_d F_{2}^{u}(t)+e_u F_{2}^{d}(t).
	\label{eq:n2}
\end{equation}

where $e_u = 2/3$ and $e_d = -1/3$ are the corresponding quark electric charges~\cite{Diehl:2013xca}.

\begin{table}[H]
	\begin{center}
		\caption{{\footnotesize Coefficients for calculations of the proton  form factors, Group 1, ($tF_1^{p}$, $tF_2^{p}$, $G_E^{p}$, and $G_M^{p}$). The fit was performed simultaneously for all four form factors of the proton, with parameters $b$ and $m'$ fixed from Refs.\cite{HajiHosseiniMojeni:2022tzn,Vaziri:2023xee}, and the remaining parameters obtained from the data~\cite{Qattan:2012zf}.}
			\label{tab:Proton}}
		\vspace*{0.4mm}
		\begin{tabular}{l@{\hskip 7mm}c}
			\hline\hline
			\\[-1.5ex]
			\multicolumn{2}{c}{\textbf{Combined Model For Proton (Group 1)}~\cite{Khorramian:2009xz,Martin:2002dr,Vaziri:2024fud,Guidal:2004nd}} \\
			\hline
			\\[-1.5ex]
			$\alpha^{\prime\prime}$ & $2.067 \pm 0.098$ \\
			$\beta$                 & $2.854 \pm 0.122$  \\
			$\gamma$                & $0.010 \pm 0.013$  \\
			$W_1$                   & $0.035 \pm 0.003$ \\
			$W_2$                   & $0.040 \pm 0.008$ \\
			$W_3$                   & $0.925 \pm 0.008$  \\
			$b$                     & $2.00$ (fixed)        \\
			$m^{\prime}$            & $0.65$ (fixed)     \\
			\hline
			$\chi^2 / n$ (\textbf{Total}) & $1.67$,  \, $n=141$ \\
			\hline\hline
		\end{tabular}
	\end{center}
\end{table}

For the Sachs form factors we employ the standard relations already given in Eqs.~(\ref{eq:sachs1}--\ref{eq:sachs2}) of Sec.~\ref{sec:sec2}.

\begin{table}[H]
	\begin{center}
		\caption{{\footnotesize Coefficients for the combined model of $G_M^p$, $G_E^p$, and $G_M^n$, Group 2, obtained from a simultaneous fit. Parameters $b$ and $m^{\prime}$ were fixed fixed from Refs.\cite{HajiHosseiniMojeni:2022tzn,Vaziri:2023xee}, while the others were fitted.}
			\label{tab:GGG}}
		\vspace*{0.4mm}
		\begin{tabular}{l@{\hskip 7mm}c}
			\hline\hline
			\\[-1.5ex]
			\multicolumn{2}{c}{\textbf{Combined Model: $G_M^{P}$, $G_E^{p}$, $G_M^{n}$} (Group 2)~\cite{Khorramian:2009xz,Martin:2002dr,Vaziri:2024fud,Guidal:2004nd}~} \\
			\hline
			\\[-1.5ex]
			$\alpha^{\prime\prime}$ & $0.947 \pm  0.071$ \\
			$\beta$                 & $0.100 \pm  0.294$  \\
			$\gamma$                &  $1.409 \pm  0.175$  \\
			$W_1$                   & $0.049 \pm0.011$ \\
			$W_2$                   & $0.153 \pm0.035$ \\
			$W_3$                   & $0.799 \pm 0.028$  \\
			$b$                     & $2.00$ (fixed)        \\
			$m^{\prime}$            & $0.65$ (fixed)     \\
			\hline
			$\chi^2 / n$ (\textbf{Total}) & $1.16$ ,  \, $n=117$ \\
			\hline\hline
		\end{tabular}
	\end{center}
\end{table}

\begin{figure*}
	%\vspace*{-.658cm}
	\includegraphics[clip,width=0.45\textwidth]{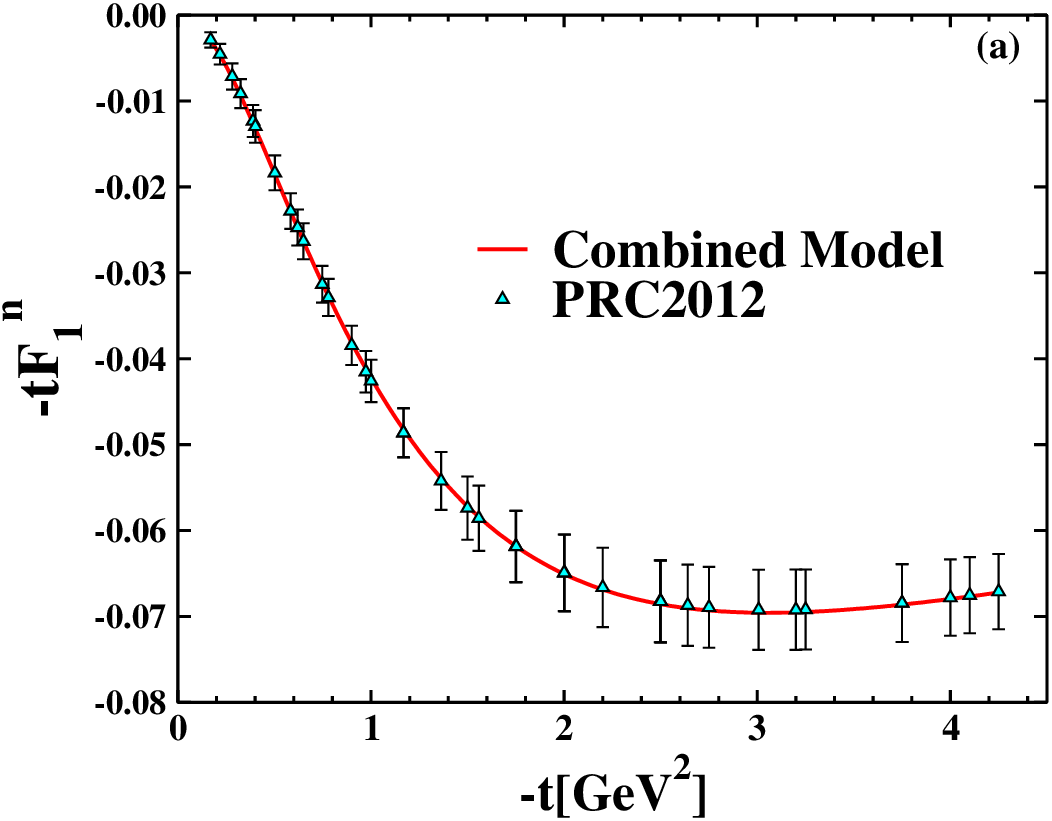}
	\hspace*{2mm}
	\includegraphics[clip,width=0.45\textwidth]{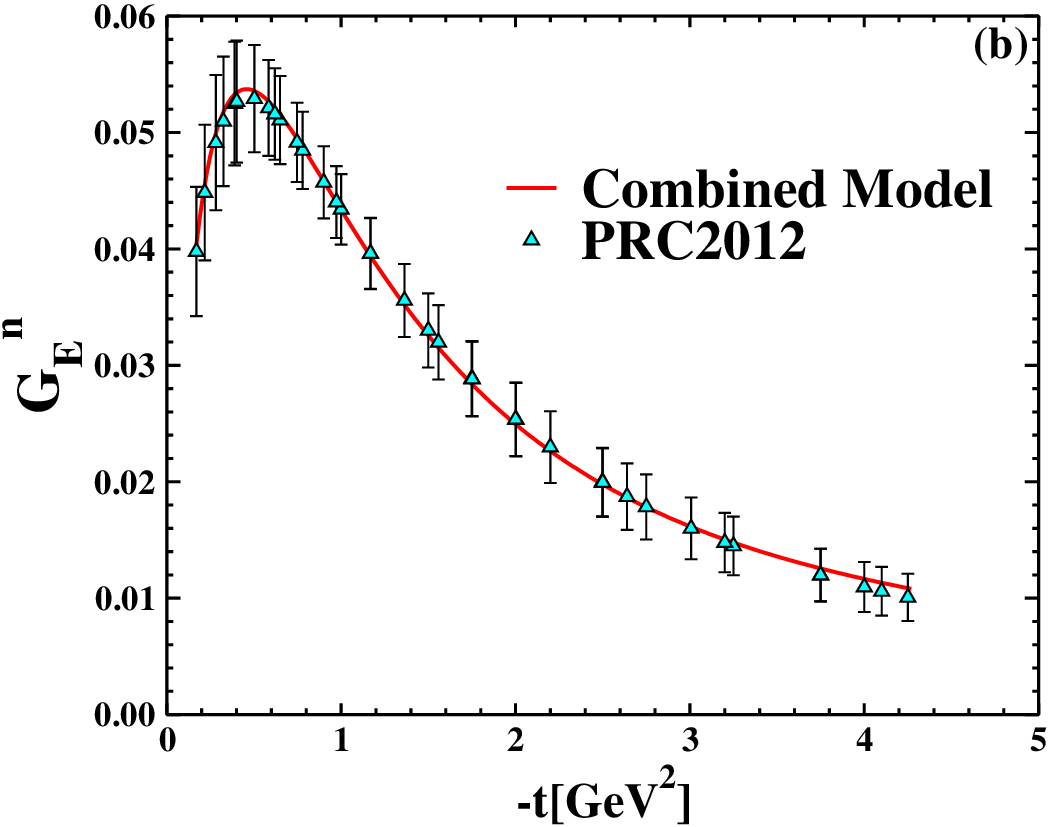}
	
	\	\caption{\footnotesize  The combined model of  $tF_1^{n}$ and $G_E^{n}$ ((group 4)) as a function of $-t$. The ER Ansatz~\cite{Guidal:2004nd}+MRST2002 PDFs~\cite{Martin:2002dr}, the VS24 Ansatz~\cite{Vaziri:2024fud} with new parameters+ new PDF(Eq.~\eqref{eq:newpdf}), and the VMD model~\cite{Masjuan:2012sk} are used.We fit Combined model to the nucleon form factors, whose numerical values were extracted from experimental data and reported by the authors in Ref.~\cite{Qattan:2012zf}.}
	\label{fig:neutron2}
\end{figure*}

	\section{Combination of Three Models}\label{sec:sec4}
	Using different approaches, we fit the nucleon form factors to those extracted from electron-nucleon scattering data. As introduced in Secs.~(\ref{sec:sec2}--\ref{sec:sec3}), three functional models are employed to describe the nucleon form factors. Two of them are based on the GPD framework and are defined in Sec.~(\ref{sec:sec3}): the ER ansatz~\cite{Guidal:2004nd} combined with the MRST2002 PDF set~\cite{Martin:2002dr}, and the VS24 ansatz~\cite{Vaziri:2024fud} combined with the KKA10 PDF set~\cite{Khorramian:2009xz}. The third one is based on the VMD model~\cite{Masjuan:2012sk} described in Sec. (\ref{sec:sec2}). Since the individual use of each model did not simultaneously reproduce all nucleon form factors in agreement with experimental data, a combined approach is expected to yield more consistent and insightful results. 
	Then, we  assign a weighting factor to each of the three functions, $W_1$, $W_2$, and $W_3$ under the condition that their sum equals unity. By fitting the combined expression to the experimental data ~\cite{Qattan:2012zf}, 4we obtained the parameters, including the weights $W_{1}, W_{2}, W_{3}$ and the VS24 shape parameters $\alpha'', \beta, \gamma$. The expressions fitted are:
	
   \begin{equation}
   	\begin{aligned}
   		F_{1,2}^{(p,n)}|_{\text{combined model}}(t) =\;& 
   		W_1\, F_{1,2}^{(p,n)}|_{\text{VMD}}(t)\\ + 
   		W_2\, F_{1,2}^{(p,n)}|_{\text{VS24}}(t) 
   		& + 
   		W_3\, F_{1,2}^{(p,n)}|_{\text{ER}}(t),
   	\end{aligned}
   	\label{eq:F12 combined}
   \end{equation}

Similarly,
\begin{equation}
	\begin{aligned}
		G_{M,E}^{(p,n)}|_{\text{combined model}}(t) =\;& 
		W_1\, G_{M,E}^{(p,n)}|_{\text{VMD}}(t)\\ + 
		W_2\, G_{M,E}^{(p,n)}|_{\text{VS24}}(t) 
		& + 
		W_3\, G_{M,E}^{(p,n)}|_{\text{ER}}(t),
	\end{aligned}
	\label{eq:GMGEcombined}
\end{equation}

The fitting procedure was carried out in four separate stages for four groups of form factors, which we labeled as Groups $1$ to $4$. First, we determined the six parameters of $F_{1,2}^{(p,n)}|_{\text{combined model}}(t)$ and $G_{M,E}^{(p,n)}|_{\text{combined model}}(t)$ for the proton by fitting them simultaneously to the experimental data~\cite{Qattan:2012zf}, where the coefficients $b$ and $m'$ were fixed from Refs.\cite{HajiHosseiniMojeni:2022tzn,Vaziri:2023xee}. Letting them free would introduce extra correlations without returning new information into the fits. The resulting form factors are plotted as functions of $-t$ in Fig.~\ref{fig:Proton}, and the extracted parameter values are summarized in Tab.~\ref{tab:Proton} for Group $1$.

The next step involved calculating the electromagnetic form factors $G_E^{p}$, $G_M^{p}$, and $G_M^{n}$, Group 2. The corresponding results are displayed in Fig.~\ref{fig:GGG}, and the fitted coefficients are summarized in Tab.~\ref{tab:GGG}. The three form factors were fitted simultaneously. 

 Since the internal structure of the neutron is more complex than that of the proton, the neutron fits were carried out in two separate stages.
 We first fitted the functions of \( tF_{2}^n \) and \( G_{M}^n \), Group 3, to the experimental data~\cite{Qattan:2012zf}, simultaneously.  Fig.~\ref{fig:neutron1} and  Tab. \ref{tab:neutron1} show the results of these calculations.  In particular, calculating  two form factors \( tF_{1}^n \) and \( G_{M}^n \) requires a larger set of fit parameters. In this analysis, in addition to the parameters introduced in the previous sections, we also fitted the following parton distribution parameters~\cite{Diehl:2013xca,Mojeni:2020rev} :

\begin{equation}
	\begin{aligned}
		x u_v(x) &= A_u\, x^{a_u} (1 - x)^{b_u} \left(1 + p_u\, \sqrt{x} + \gamma_u\, x \right), \\
		x d_v(x) &= A_d\, x^{a_d} (1 - x)^{b_d} \left(1 + p_d\,\sqrt{x} + \gamma_d\, x \right),
	\end{aligned}
	\label{eq:newpdf}
\end{equation}
which we refer to as a new PDF parametrization, where the normalization constants \( A_u \) and \( A_d \) are given by

\begin{equation}
	\begin{aligned}
		A_u &= \dfrac{2}{
			\begin{aligned}
				&B(a_u, b_u + 1) + p_u\, B(a_u + \tfrac{1}{2}, b_u + 1) \\
				&\quad + \gamma_u\, B(a_u + 1, b_u + 1)
			\end{aligned}
		}, \\
		A_d &= \dfrac{1}{
			\begin{aligned}
				&B(a_d, b_d + 1) + p_d\, B(a_d + \tfrac{1}{2}, b_d + 1) \\
				&\quad + \gamma_d\, B(a_d + 1, b_d + 1)
			\end{aligned}
		}.
	\end{aligned}
\end{equation}

Here, \( B(a, b) \) denotes the Euler beta function, defined as~\cite{Martin:2002dr,Abramowitz:1972}
\[
B(a, b) = \int_0^1 x^{a-1} (1 - x)^{b-1} \, dx.\]

The graphs of  \( tF_{1}^n \) and \( G_{E}^n \) as function of $-t$ were shown in Fig.~\ref{fig:neutron2}, and the parameters obtained from fitting these form factors are listed in Tab. \ref{tab:neutron2} for Group $4$. These two form factors were fitted simultaneously. The small value of  $\chi^2 / ndf = 0.25$, with $ndf$ the number of degrees of freedom, reflects the relatively large experimental uncertainties and the limited number of data points in this channel, together with a clear degree of overfitting. This is reflected by the fact that key parameters such as $W_3$, $\gamma_{u,d}$, $\alpha''$, are clearly well determined while subleading ones such as $\beta$, $\gamma$, or $p_{u,d}$ suffer large uncertainties.

In all our calculations, the values of $\eta_u$ for the ER~\cite{Guidal:2004nd} and VS24 ansatzes~\cite{Vaziri:2024fud} are $1.713$ and $0.69$, respectively, while the corresponding $\eta_d$ values are $0.566$ and $0.27$.

    \begin{table}[H]
    	\begin{center}
    		\caption{{\footnotesize Coefficients for calculations of form factors for the  $tF_2^{n}$ and $G_M^{n}$, Group 3, obtained from a simultaneous fit. The parameters $b$ and $m^{\prime}$ are fixed; the others have been calculated by fitting.}
    			\label{tab:neutron1}}
    		\vspace*{0.4mm}
    		\begin{tabular}{l@{\hskip 7mm}c}
    			\hline\hline
    			\\[-1.5ex]
    			\multicolumn{2}{c}{\textbf{Combined Model For Neutron (Group 3)}~\cite{Khorramian:2009xz,Martin:2002dr,Vaziri:2024fud,Guidal:2004nd}} \\
    			\hline
    			\\[-1.5ex]
    			$\alpha^{\prime\prime}$ & $1.050 \pm  0.004$ \\
    			$\beta$                 & $1.000 \pm 0.003$  \\
    			$\gamma$                &  $0.005 \pm 0.002$  \\
    			$W_1$                   & $0.050 \pm 0.002$ \\
    			$W_2$                   & $0.394 \pm 0.004$ \\
    			$W_3$                   & $0.556 \pm 0.004$  \\
    			$b$                     & $2.00$ (fixed)        \\
    			$m^{\prime}$            & $0.65$ (fixed)     \\
    			\hline
    			$\chi^2 / ndf$   & $1.31$ ($ndf=78$)\\
    			\hline\hline
    		\end{tabular}
    	\end{center}
    \end{table}

\begin{table}[H]
	\begin{center}
		\caption{{\footnotesize Coefficients for calculations of form factors for the  $tF_1^{n}$,  and $G_E^{n}$, Group 4, obtained from a simultaneous fit. The parameters $b$ and $m^{\prime}$ are fixed; the others have been calculated by fitting.}
			\label{tab:neutron2}}
		\vspace*{0.4mm}
		\begin{tabular}{l@{\hskip 7mm}c}
			\hline\hline
			\\[-1.5ex]
			\multicolumn{2}{c}{\textbf{Combined Model: $tF_1^{n}$ and $G_E^{n}$ (Group 4)}~\cite{Khorramian:2009xz,Martin:2002dr,Vaziri:2024fud,Guidal:2004nd}} \\
			\hline
			\\[-1.5ex]
			$\alpha^{\prime\prime}$ & $0.711 \pm 0.084$ \\
			$\beta$                 & $-0.290 \pm 0.093$  \\
			$\gamma$                &  $0.111 \pm 0.101$  \\
			$\alpha$                &  $1.150 \pm 0.012$  \\
			$a_u$                   & $0.220 \pm 0.074$  \\
			$b_u$                   & $0.629 \pm 0.1384$  \\
			$p_u$                   & $0.073 \pm 3.623$  \\
			$\gamma_u$              & $44.444 \pm 6.051$  \\
			$a_d$                   & $3.466 \pm 0.624$  \\
			$b_d$                   & $19.540 \pm 2.655$  \\
			$p_d$                   & $10.294 \pm 2.909$  \\
			$\gamma_d$              & $15.772 \pm 1.203$  \\
			$W_1$                   & $0.126 \pm 0.117$ \\
			$W_2$                   & $0.167 \pm 0.030$ \\
			$W_3$                   & $0.716 \pm 0.009$  \\
			$b$                     & $2.00$ (fixed)        \\
			$m^{\prime}$            & $0.65$ (fixed)     \\
			\hline

			$\chi^2 / ndf$ & $0.25$ ($ndf=56$) \\
			
			\hline\hline
		\end{tabular}
	\end{center}
\end{table}

\vspace{3mm}
\begin{table}[H]
	\centering
	\scriptsize
	\caption{Effective contribution range of each component in the combined model across different $Q^2$ regions.}
	\label{tab:Q2Ranges}
	\begin{tabular}{l@{\hskip 0.1mm}c}
		\noalign{\hrule height 1.5pt}  % ?? ????? ??? ?? ??? ???
		\textbf{$Q^2$ Range} & $Q^2 < 1\, \mathrm{GeV}^2$ \\
		\hline
		\textbf{Dominant Model} & VMD \\
		\hline
		\textbf{Reason for Dominance} & \\- Strong vector meson exchange; \\  - low-energy regime. \\
		\hline
		\textbf{Component} & $W_3$ \\\\
		\noalign{\hrule height 1.5pt}  % ?? ????? ??? ?? ??? ???
		\hline
		\textbf{$Q^2$ Range} & $1 < Q^2 < 3\, \mathrm{GeV}^2$ \\
		\hline
		\textbf{Dominant Model} & VS24 + KKA10 \\
		\hline
		\textbf{Reason for Dominance} &\\ - Flexible exponential tail \\
		 - PDF sensitivity. \\
		\hline
		\textbf{Component} & $W_2$ \\\\
		\noalign{\hrule height 1.5pt}  % ?? ????? ??? ?? ??? ???
		\hline
		\textbf{$Q^2$ Range} & $Q^2 > 3\, \mathrm{GeV}^2$ \\
		\hline
		\textbf{Dominant Model} & ER + MRST2002 \\
		\hline
		\textbf{Reason for Dominance} & \\ -Sharp fall-off; $x\to 1$ behavior in GPDs \\
		\hline
		\textbf{Component} & $W_1$ \\
		\hline
	\end{tabular}
\end{table}
\vspace{2mm}

The quality of the fits is quantified by the $\chi^2/ndf$ values reported for each observable and for the combined fit. Uncertainties are treated through a standard $\chi^2$ minimization, using the experimental errors as weights, while parameter uncertainties and correlations are obtained from the Minuit covariance matrix after convergence~\cite{Minuit2}.

It is evident from the figures that the form factors obtained within the combined framework of the three models are in excellent agreement with those extracted from electron--proton scattering experiments~\cite{Qattan:2012zf}.

The effective contribution ranges of the individual components of the combined model across different $Q^2$ regions are summarized in Tab.~\ref{tab:Q2Ranges}. This table provides a clear partitioning of the kinematic domains in which each model component plays a dominant role, reflecting the underlying physical mechanisms governing different momentum-transfer regimes as obtained from the fits.

At low momentum transfers, $Q^2 < 1~\mathrm{GeV}^2$, the \textbf{VMD} component dominates the description. This behavior is physically expected, as strong vector-meson exchange mechanisms are known to govern the low-energy regime. In this region, the contribution associated with the weight factor $W_3$ is found to be dominant. This is also the region where isospin breaking effects, and the half-width rule have less effect.

In the intermediate region, $1 < Q^2 < 3~\mathrm{GeV}^2$, the description is primarily driven by the combined \textbf{VS24 + KKA10} components, e.g. parton distribution functions at $N^3LO$. This domain benefits from the flexible exponential behavior and the sensitivity to PDFs inherent in these models, with the corresponding contribution governed by the weight factor $W_2$.

At higher momentum transfers, $Q^2 > 3~\mathrm{GeV}^2$, the \textbf{ER + MRST2002} framework provides the dominant contribution. This is specially so for the Regge ER model use as this region is characterized by a steep fall-off of the form factors, closely related to the $x \to 1$ behavior of generalized parton distributions (GPDs). Consequently, the contribution associated with the weight factor $W_1$ becomes dominant.

Overall, the structure summarized in Tab.~\ref{tab:Q2Ranges} demonstrates that the combined model consistently incorporates the relevant physics across a wide range of momentum transfers, assigning dominance to each component precisely in the kinematic region where its underlying physical interpretation is most appropriate.

\section{Phenomenological Representations of Neutron Form Factors}\label{sec:sec5}

Before analyzing the results and performing numerical investigations, inspired by the $Q^2$-region results from the previous section and to gain deeper insight into the internal structure of the neutron, we explore two complementary phenomenological representations of its electromagnetic form factors in the intermediate momentum transfer region. Specifically, we analyze the scaled Dirac form factor, \( n(t) = t \cdot F_1^n(t) \), and the magnetic form factor, \( G_M^n(t) \), using a Regge-inspired linear trajectory and a Pad\'e-like rational function fit~\cite{Collins:1977jy,Donnachie:2002en,Arrington:2007ux,Masjuan:2012wy,Escribano:2015nra,Escribano:2015yup}. These are the most difficult form factors to extract from data as data are scarce and, as such, we propose to simplify the models to a mininum yet significant set up.

For the Dirac component, this function is expected to exhibit approximate linear behavior in analogy with Regge trajectories, commonly used to describe the high-energy asymptotics of hadronic amplitudes~\cite{Collins:1977jy, Donnachie:2002en}. The model assumes a linear form:
\begin{equation}
	\label{eq:regge_linear}
	n(t) = \alpha_0 + \alpha' t,
\end{equation}
with fitted parameters:
\begin{equation}
	\alpha_0 = 0.028, \quad \alpha' = -0.085~\mathrm{GeV}^{-2}.
\end{equation}
This slope \( \alpha' \) is significantly smaller than typical baryonic Regge slopes (\( \sim 0.9 \, \mathrm{GeV}^{-2} \)), indicating that the fitted function captures soft physics at moderate $Q^2$ without requiring a steep Regge rise. 

The neutron magnetic form factor \( G_M^n(t) \) is fitted using the Pad\'e-like function~\cite{Arrington:2007ux,Masjuan:2012wy,Escribano:2015nra,Escribano:2015yup}:
\begin{equation}
	G_M^n(t) = \frac{a + b t}{c + d t + e t^2 + f t^3}.
\end{equation}
This functional form provides a flexible and stable interpolation across the available data, while preserving the correct low-$t$ behavior and suppressing unphysical growth at larger $t$. 

\begin{figure}[H]
	\centering
	\includegraphics[width=0.45\textwidth]{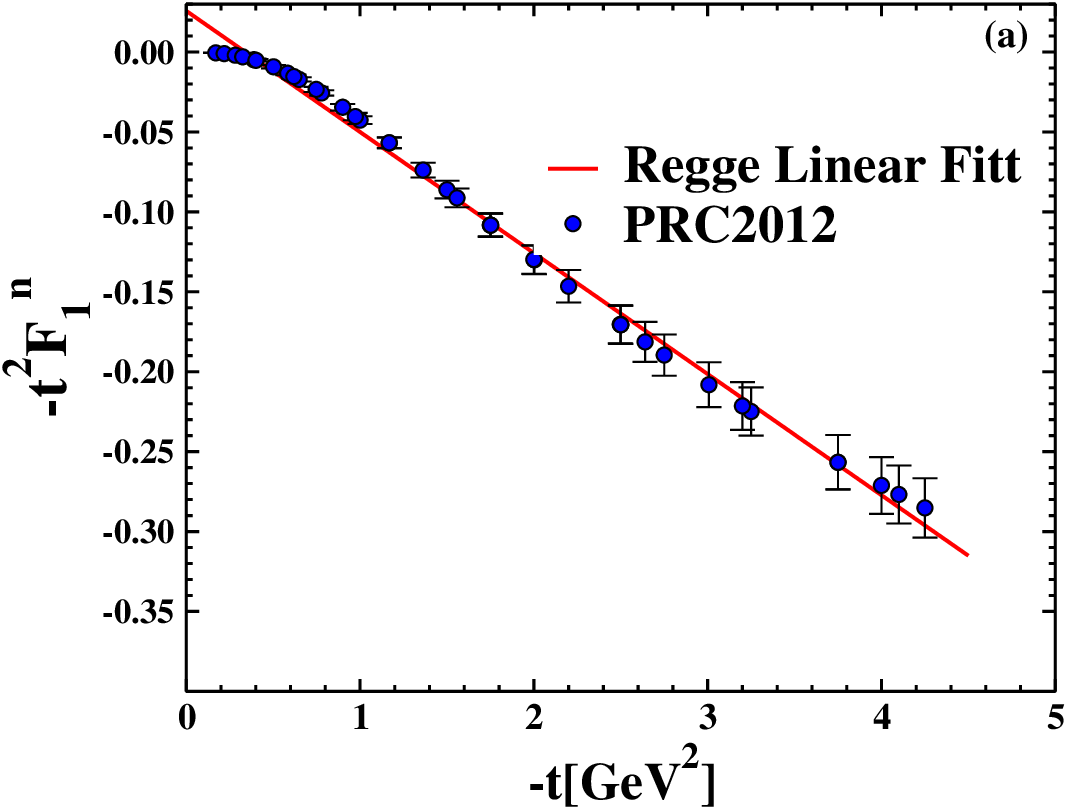}
	\hspace*{2mm}
	\includegraphics[clip,width=0.45\textwidth]{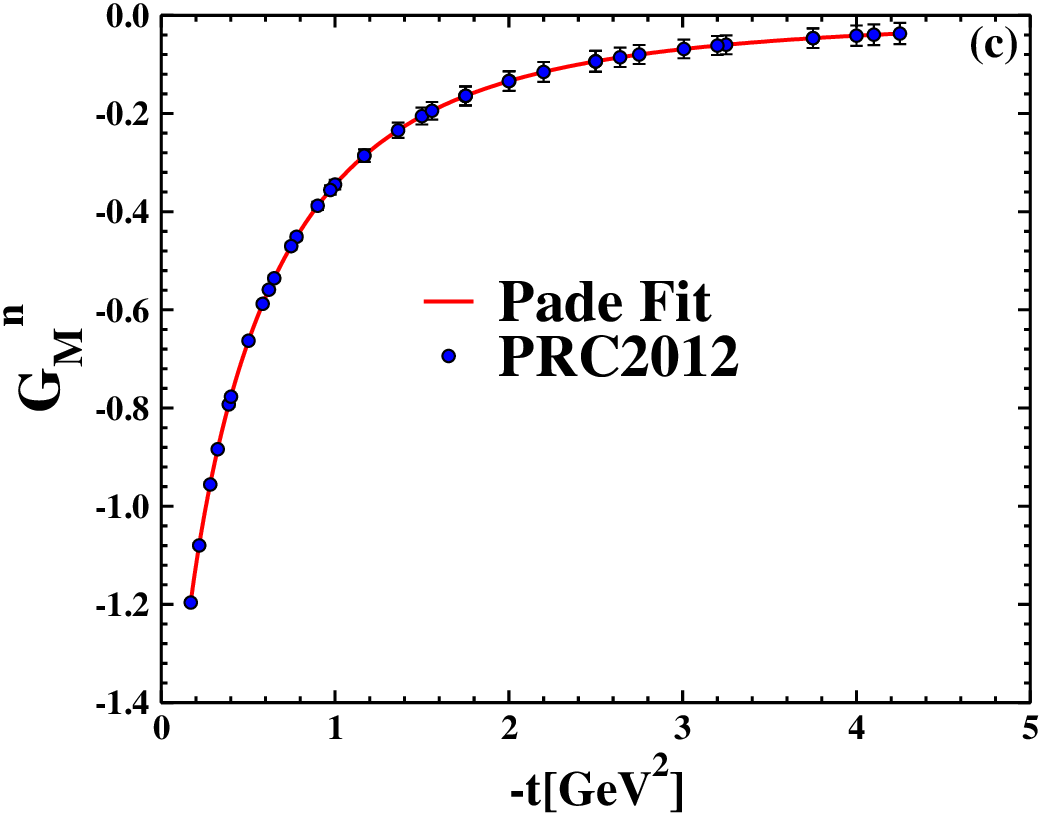}
	\vspace*{3mm}\\
	\caption{Comparison of phenomenological fits with neutron form factor data: (a) Regge-inspired linear~\cite{Collins:1977jy, Donnachie:2002en} fit for \( tF_1^n \), (b) Pad\'e~\cite{Arrington:2007ux,Masjuan:2012wy,Escribano:2015nra,Escribano:2015yup} fit for \( G_M^n \) using data from Ref.~\cite{Qattan:2012zf}.}
	\label{fig:tf1nterajectory}
\end{figure}

As shown in Fig.~\ref{fig:tf1nterajectory}, the Regge trajectory and Pad\'e approximants fits closely follow the experimental data~\cite{Qattan:2012zf} and have very small fitting errors, making them suitable for precise applications and natural candidates for use in the global fitting procedure introduced in the next section.

These phenomenological representations serve as a validation step for the combined models and provide stable analytic forms that will be incorporated into the subsequent global fits.

\section{Numerical Analysis using Pad\'e approximants} \label{sec:sec6}

In this final section, we attempt to collect the physical insights gathered in Table \ref{tab:Q2Ranges}, i.e., how the different physical intuition in the construction of the different models arise in the different fits. The first thing to realize is that when combining models into a fit, the different models overlap, as is notoriously seen by the fact that weighting factors $W_{1,2,3}$ appear in all fits. The second thing is that, for practice purposes, it would be wiser to have a unique function describing the different form factors. Our proposal is, then, to use a fully general parameterization based in Pad\'e approximants (PA)\cite{Masjuan:2012wy,Escribano:2015nra,Escribano:2015yup} to describe the form factors while reporting the coefficients of PAs for each of the four grups of FFs. This procedure ensures that each form factor is represented by a smooth and stable analytic form, providing well-behaved coefficients for the Pad\'e construction and reducing extrapolation instabilities. The insights from the Regge trajectory analysis also guide the expected moderate- and high-\(t\) behavior, making the resulting Pad\'e parametrizations physically consistent with known QCD scaling trends.

\begin{figure*}
	\includegraphics[clip,width=0.45\textwidth]{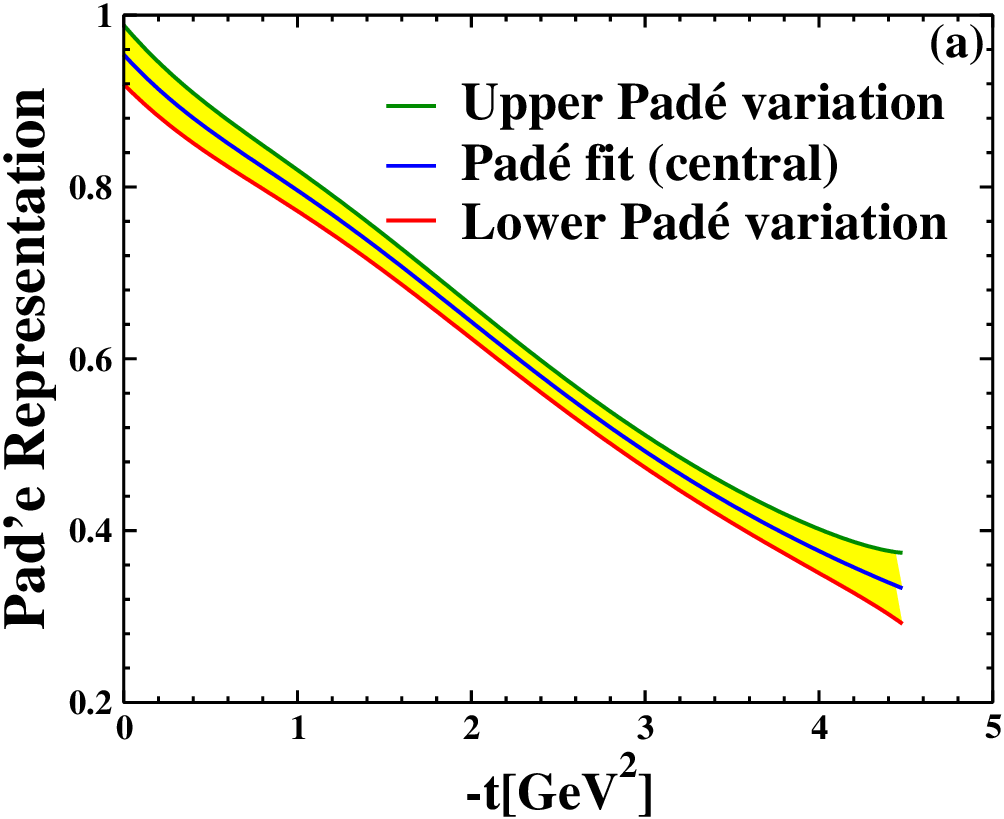}
	\includegraphics[clip,width=0.45\textwidth]{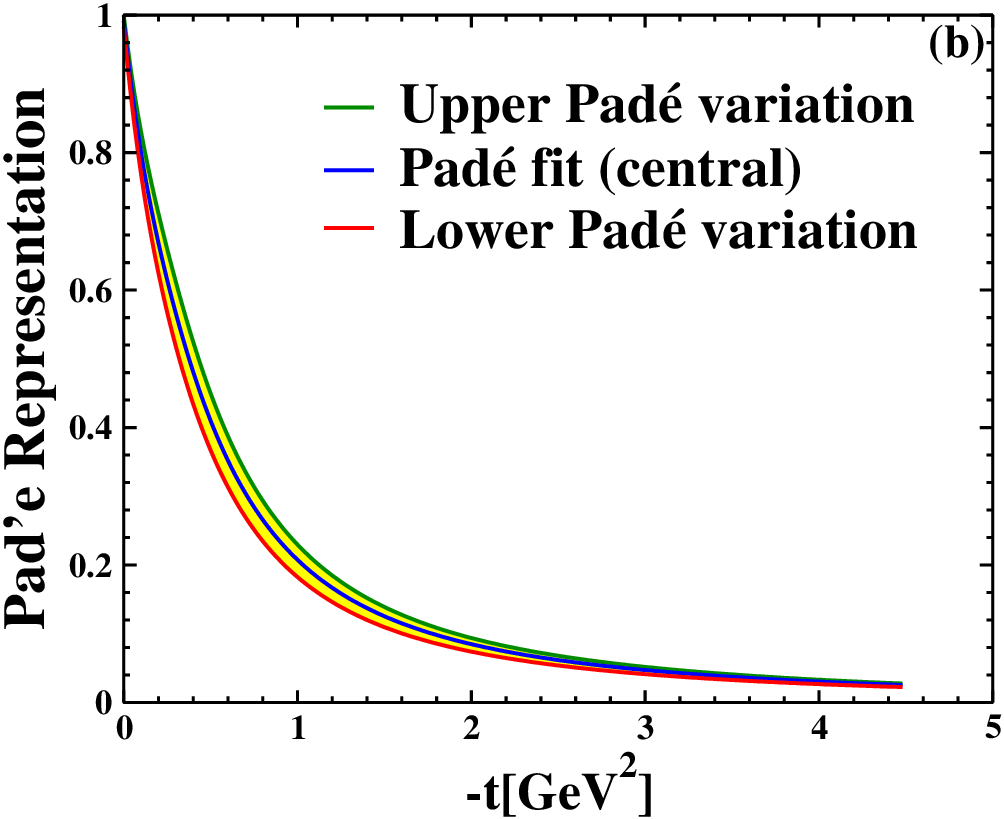}
	\includegraphics[clip,width=0.45\textwidth]{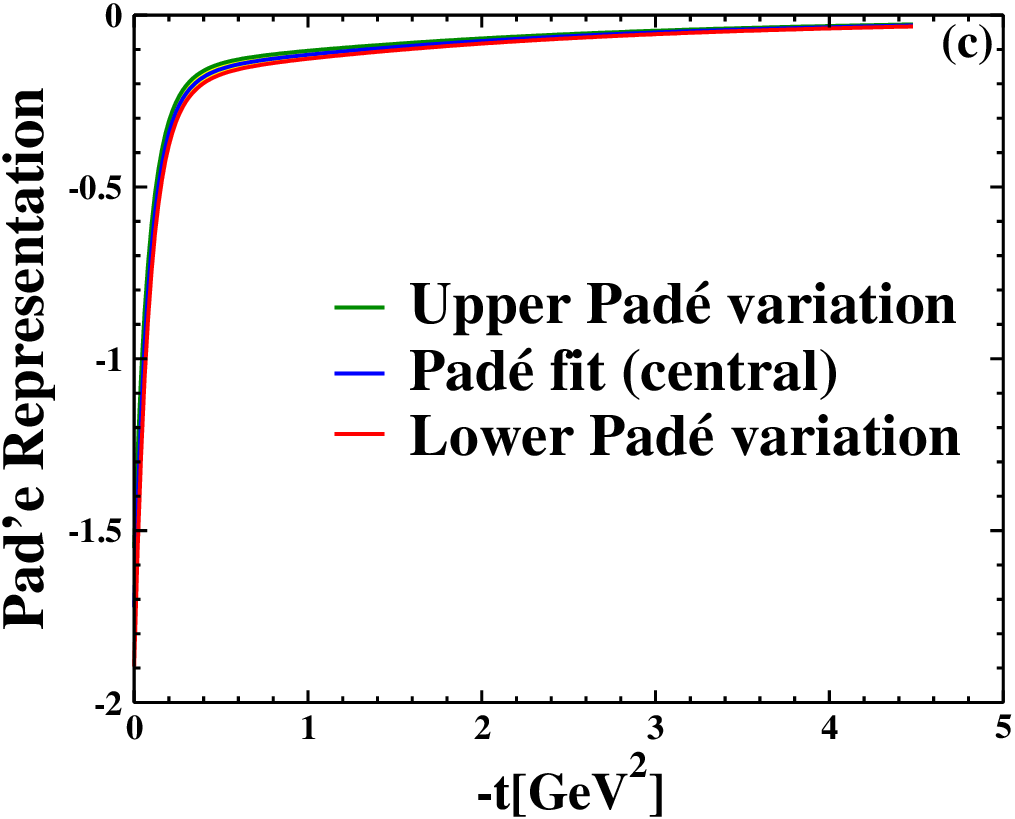}
	\includegraphics[clip,width=0.45\textwidth]{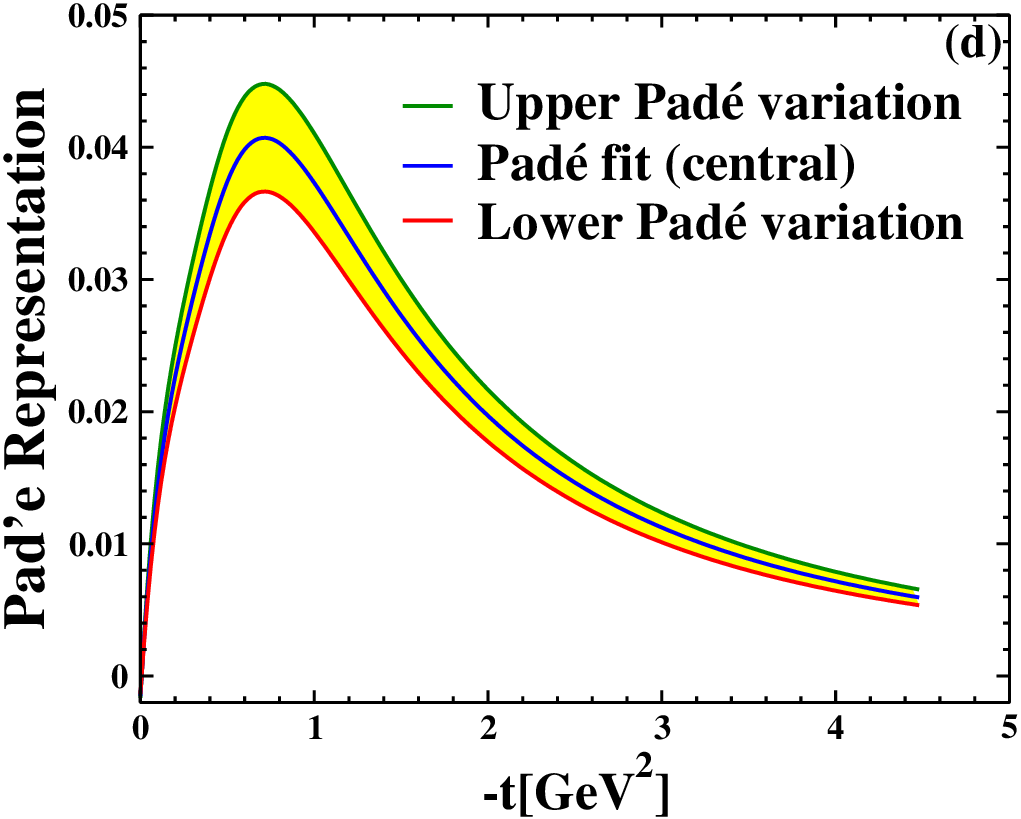}
	\caption{\footnotesize
		Pad\'e parametrizations for four groups of form factors as functions of $-t$  :  
		(a) $tF_1^{p}$, $tF_2^{p}$, $G_E^{p}$, and $G_M^{p}$ (group 1);  
		(b) $G_E^{p}$, $G_M^{p}$, and $G_M^{n}$ (group2);  
		(c) $tF_2^{n}$ and $G_M^{n}$ (group 3);  
		(d) $tF_1^{n}$ and $G_E^{n}$ (group 4).  
		The Pad\'e representation for these parametrizations is given in Eq.~(\ref{eq:pade_general}), and the corresponding parameters are listed in Tabs.~\ref{tab:group1}--\ref{tab:group4}.The shaded bands indicate a model-dependent sensitivity estimate of the Pad\'e parametrization,
		obtained from controlled variations of the Pad\'e coefficients as described in the text.They should not be interpreted as statistical error bands.}
	\label{fig:GlobalFits}
\end{figure*}

The workflow proceeds as follows: (i) local fits (polynomial/Taylor expansions) were performed to extract the numerical behavior of each form factor; (ii) Pad\'e approximants were constructed from those coefficients in order to obtain compact rational representations; (iii) Pad\'e parametrizations were then constructed for each group and compared with experimental data~\cite{Qattan:2012zf}. This strategy follows common practice in parametrizations of hadronic observables~\cite{Hohler:1976ax,Martin:2009iq}. We employ the standard Pad\'e form
\begin{equation}
	P_{[N,M]}(t) \;=\; \frac{a_0 + a_1 t + a_2 t^2 + \cdots + a_N t^N}
	{1 + b_1 t + b_2 t^2 + \cdots + b_M t^M},
	\label{eq:pade_general}
\end{equation}
where the numerator degree \(N\) and denominator degree \(M\) are chosen case-by-case to balance accuracy in the low-\(t\) region and stability at larger \(t\). The coefficients \(a_i\) and \(b_j\) for four groups of form factors are reported in Tabs.~\ref{tab:group1}--\ref{tab:group4}. The resulting Pad\'e parametrizations, together with the experimental data points, are displayed in Fig.~\ref{fig:GlobalFits}.
As can be seen, the resulting parametrizations exhibit the expected qualitative behavior in the limits $t \to 0$ and $t \to \infty$, consistent with QCD expectations.

% ---------------- Table: Group 1 (4 form factors) ----------------
\begin{table}[H]
	\begin{center}
		\caption{{\footnotesize Pad\'e coefficients for group 1. The Pad\'e form is given in Eq.~\eqref{eq:pade_general}. The corresponding Pad\'e parametrization for this group is shown in Fig.~\ref{fig:GlobalFits}(a).}}
		\label{tab:group1}
		\vspace*{0.4mm}
		\begin{tabular}{l@{\hskip 6mm}cccc}
			\hline\hline\\[-1.5ex]
			& $tF_{1}^{p}$ & $tF_{2}^{p}$ & $G_{E}^{p}$ & $G_{M}^{p}$ \\
			\hline\\[-1.5ex]
			$a_0$ & $7.33\times10^{-10}$ & $3.39\times10^{-10}$ & $1.01$ & $2.8$ \\
			$a_1$ & $1.01$ & $1.8$ & $-0.75$ & $28.3$ \\
			$a_2$ & $11.6$ & $24.5$ & $8.3$ & $120.5$ \\
			$a_3$ & $66.6$ & $160.2$ & --- & $765.7$ \\
			$a_4$ & $136.8$ & $373.7$ & --- & --- \\
			\midrule
			$b_1$ & $1$ & $1$ & $1$ & $1$ \\
			$b_2$ & $14.2$ & $16.8$ & $2.4$ & $13.1$ \\
			$b_3$ & $96.86$ & $134.7$ & $7.8$ & $74.5$ \\
			$b_4$ & $314.75$ & $514.1$ & $20.7$ & $409.05$ \\
			$b_5$ & $370.48$ & $810.3$ & $25.7$ & $865.8$ \\
			$b_6$ & $118.8$ & $481.7$ & --- & $525.06$ \\
			$b_7$ & --- & $74.8$ & --- & --- \\
			\hline\hline
		\end{tabular}
	\end{center}
\end{table}

% ---------------- Table: Group 2 (3 form factors) ----------------
\begin{table}[H]
	\begin{center}
		\caption{{\footnotesize Pad\'e coefficients for group 2 (three form factors). The corresponding Pad\'e parametrization for this group is shown in Fig.~\ref{fig:GlobalFits}(b).}}
		\label{tab:group2}
		\vspace*{0.4mm}
		\begin{tabular}{l@{\hskip 6mm}ccc}
			\hline\hline\\[-1.5ex]
			& $G_{E}^{p}$ & $G_{M}^{p}$ & $G_{M}^{n}$ \\
			\hline\\[-1.5ex]
			$a_0$ & $1.009$ & $1.68$ & $-1.8$ \\
			$a_1$ & $33.3$ & $5.4$ & $-4.9$ \\
			$a_2$ & $72.7$ & $22.95$ & $-21.8$ \\
			\midrule
			$b_1$ & $1.0$ & $1.0$ & $1.0$ \\
			$b_2$ & $35.9$ & $6.08$ & $5.6$ \\
			$b_3$ & $165.4$ & $22.8$ & $21.1$ \\
			$b_4$ & $211.8$ & $39.5$ & $38.9$ \\
			$b_5$ & $108.03$ & $13.5$ & $21.6$ \\
			\hline\hline
		\end{tabular}
	\end{center}
\end{table}

% ---------------- Table: Group 3 (2 form factors) ----------------
\begin{table}[H]
	\begin{center}
		\caption{{\footnotesize Pad\'e coefficients for group 3 (two form factors). The corresponding Pad\'e parametrization for this group is shown in Fig.~\ref{fig:GlobalFits}(c).}}
		\label{tab:group3}
		\vspace*{0.4mm}
		\begin{tabular}{l@{\hskip 6mm}cc}
			\hline\hline\\[-1.5ex]
			& $tF_{2}^{n}$ & $G_{M}^{n}$ \\
			\hline\\[-1.5ex]
			$a_0$ & $-3.4\times10^{-11}$ & $-1.76$ \\
			$a_1$ & $-1.7$ & $-11.4$ \\
			$a_2$ & $-18.6$ & $-42.4$ \\
			$a_3$ & $-104.5$ & $-33.8$ \\
			$a_4$ & $-320.5$ & --- \\
			$a_5$ & $-420.4$ & --- \\
			\midrule
			$b_1$ & $1.0$ & $1.0$ \\
			$b_2$ & $13.4$ & $9.3$ \\
			$b_3$ & $91.4$ & $43.4$ \\
			$b_4$ & $307.7$ & $95.5$ \\
			$b_5$ & $863.6$ & $84.2$ \\
			$b_6$ & $1016.5$ & $43.8$ \\
			$b_7$ & $500.4$ & --- \\
			$b_8$ & $106.03$ & --- \\
			\hline\hline
		\end{tabular}
	\end{center}
\end{table}

% ---------------- Table: Group 4 (2 form factors) ----------------
\begin{table}[H]
	\begin{center}
		\caption{{\footnotesize Pad\'e coefficients for group 4 (two form factors). The corresponding Pad\'e parametrization for this group is shown in Fig.~\ref{fig:GlobalFits}(d).}}
		\label{tab:group4}
		\vspace*{0.4mm}
		\begin{tabular}{l@{\hskip 6mm}cc}
			\hline\hline\\[-1.5ex]
			& $tF_{1}^{n}$ & $G_{E}^{n}$ \\
			\hline\\[-1.5ex]
			$a_0$ & $-4.0\times10^{-5}$ & $-0.003$ \\
			$a_1$ & $-2.6\times10^{-4}$ & $0.43$ \\
			$a_2$ & $-0.12$ & $-0.89$ \\
			$a_3$ & --- & $2.4$ \\
			\midrule
			$b_1$ & $1.0$ & $1.0$ \\
			$b_2$ & $1.0$ & $1.55$ \\
			$b_3$ & $0.51$ & $3.92$ \\
			$b_4$ & $0.37$ & $9.43$ \\
			$b_5$ & --- & $15.7$ \\
			$b_6$ & --- & $14.3$ \\
			\hline\hline
		\end{tabular}
	\end{center}
\end{table}

These results demonstrate that the Pad\'e parametrizations yield numerically stable and accurate representations of the fitted form factors across the analyzed $t$-range. The shaded bands shown in the figures illustrate the sensitivity of the Pad\'e parametrization.
 The uncertainty bands shown in Fig.~\ref{fig:GlobalFits} are not statistical confidence intervals and are not derived from the covariance matrix of the Pad\'e coefficients.
Instead, they represent a model-dependent sensitivity estimate of the Pad\'e parametrization.
Starting from the central Pad\'e fits reported in Tabs.~\ref{tab:group1}--\ref{tab:group4}, the numerator and denominator coefficients were varied in a controlled manner where they are scanned using a MonteCarlo method,
leading to upper and lower Pad\'e curves.
An effective sensitivity function $\sigma(t)$ was defined as half of the difference between these two curves, and the shaded bands in Fig.~\ref{fig:GlobalFits} illustrate the resulting sensitivity of the Pad\'e representation \textit{a la} MonteCarlo. Accordingly, only central values of the Pad\'e coefficients are quoted in Tabs.~\ref{tab:group1}--\ref{tab:group4}.
The shaded bands indicate a model-dependent sensitivity estimate of the Pad\'e parametrization,
obtained from controlled variations of the Pad\'e coefficients as described in the text. 
They should therefore not be interpreted as statistical error bands or confidence intervals.
Notice as well that the poles of PAs obtained from speca-like data cannot be interpreted as physical ressonances, c.f. \cite{Masjuan:2012wy,Escribano:2015nra,Escribano:2015yup}. The proper way to use PAs to extract resonance-pole parameters is to use time-like data and impose threshold conditions as described in \cite{Duch:2025tuu}.

\section{ RESULTS AND CONCLUSION}\label{sec:conclusion}
	
In this work, we have studied two distinct approaches for computing nucleon form factors: the VMD and GPDs. After introducing these methods, we constructed a combined model by assigning weighting coefficients to three individual parameterizations--one based on meson dominance~\cite{Masjuan:2012sk} and two derived from the GPD framework~\cite{Guidal:2004nd,Martin:2002dr,Vaziri:2024fud,Khorramian:2009xz}. By fitting this combined model to experimental data~\cite{Qattan:2012zf}, we extracted the relevant parameters and presented the resulting form factors as functions of the squared momentum transfer, $t = -q^2$.

In Sec.~\ref{sec:sec2}, we investigated the electromagnetic structure of the nucleon within the framework of meson dominance, guided by the large-$N_c$ limit of QCD. This approach allows the nucleon form factors to be modeled as a superposition of vector-meson pole contributions, which becomes increasingly accurate as $N_c \to \infty$. We began by decomposing the electromagnetic current into its isoscalar and isovector components in Eq.~\eqref{eq:jem}, denoted by $J^{(0)\mu}$ and $J^{(1)\mu}$, respectively. The corresponding nucleon matrix elements in Eq.~\eqref{eq:Matrix} introduce the isoscalar and isovector Dirac and Pauli form factors, $F_{1,2}^{(0)}$ and $F_{1,2}^{(1)}$. These form factors were then related to the observable proton and neutron form factors using linear combinations in Eq.~\eqref{eq:ffpn}. Their physical normalization at zero momentum transfer was enforced via Eq.~\eqref{eq:f1f2norm}, reflecting the known electric charges and anomalous magnetic moments. To provide a direct physical interpretation, the Sachs electric and magnetic form factors were defined in Eqs.~\eqref{eq:sachs1}--\eqref{eq:sachs2}, which also encodes their normalization conditions at $t=0$. These quantities are crucial in connecting theoretical predictions with experimental data from electron scattering~\cite{Qattan:2012zf}. It justifies the need for multiple vector-meson poles in the parameterization of $F_1$ and $F_2$, as implemented in Eq.~\eqref{eq:Fmodel} using products of monopole terms for both isoscalar and isovector channels. Finally, the constants $c_0$ and $c_1$ in Eq.~\eqref{eq:Fmodel} were determined via matching conditions in Eq.~\eqref{eq:c0c1}, using known meson--nucleon and meson--photon couplings. In the combined model, this method enters with the weight parameter $W_1$.

GPDs encode the three-dimensional structure of nucleons. 
The formalism of GPDs is presented in detail in Eqs.~(\ref{eq:f0}--\ref{eq:EER}) in Sec.~\ref{sec:sec3}. 
The PDFs and ansatzes used in this work are summarized in Eqs.~(\ref{eq:HER}--\ref{eq:7}). 
The Pauli and Dirac form factors of the nucleons, $F_{1,2}^{p,n}$, are obtained from Eqs.~(\ref{eq:p1}--\ref{eq:n2}). 
The Sachs form factors are then used to derive the electromagnetic form factors by Eq.~(\ref{eq:sachs1}--\ref{eq:sachs2}).

Since none of these models individually met our expectations for simultaneously reproducing all form factors in agreement with experimental data, Sec.~\ref{sec:sec4} presents the core analysis of this study. In this section, combined models for four groups are constructed by assigning weight parameters $W_{1,2,3}$ to the VMD~\cite{Masjuan:2012sk}, VS24+KKA10~\cite{Vaziri:2024fud,Khorramian:2009xz}, and ER+MRST2002~\cite{Guidal:2004nd,Martin:2002dr} components, respectively.

 The combined expressions are defined in Eqs.~(\ref{eq:F12 combined}--\ref{eq:GMGEcombined}) and fitted to experimental data~\cite{Qattan:2012zf}. 
 The results for the proton are summarized in Tab.~\ref{tab:Proton} and Fig.~\ref{fig:Proton}, while the combined fits for $G_E^p$, $G_M^p$, and $G_M^n$ are shown in Tab.~\ref{tab:GGG} and Fig.~\ref{fig:GGG}. 
 The neutron analysis required a two-step procedure: the results for $tF_2^n$ and $G_M^n$ are presented in Tab.~\ref{tab:neutron1} and Fig.~\ref{fig:neutron1}, and those for $tF_1^n$ and $G_E^n$ including the new PDF of Eq.~\eqref{eq:newpdf} are given in Tab.~\ref{tab:neutron2} and Fig.~\ref{fig:neutron2}. 
 Finally, the effective ranges of each component across different $Q^2$ domains are summarized in Tab.~\ref{tab:Q2Ranges}.
 
Sec.~\ref{sec:sec5} examined two phenomenological representations of the neutron form factors as a validation step for the combined model. 
A Regge-inspired linear fit was applied to the scaled Dirac form factor $tF_1^n(t)$~\cite{Collins:1977jy,Donnachie:2002en}, yielding a slope $\alpha'=-0.085~\mathrm{GeV}^{-2}$, which is much smaller than typical baryonic Regge slopes and indicates a milder $t$-dependence at moderate momentum transfer. 
In addition, a Pad\'e-like parametrization~\cite{Arrington:2007ux} for $G_M^n(t)$ provided an accurate description of the experimental data~\cite{Qattan:2012zf}, as shown in Fig.~\ref{fig:tf1nterajectory}. 
These analytic representations reproduce the measured neutron form factors with high precision and will serve as stable inputs for the global fits developed in the next section.

Sec.~\ref{sec:sec6} presented the numerical construction of Pad\'e approximants for the nucleon form factors.
Starting from local polynomial expansions, these were extended to compact rational Pad\'e parametrizations that provide smooth and stable analytic representations over the explored $t$ range.
The extracted Pad\'e coefficients for the four groups of form factors studied in Sec.~\ref{sec:sec4} are reported in Tabs.~\ref{tab:group1}--\ref{tab:group4}, while the corresponding Pad\'e parametrizations, together with the experimental data points, are displayed in Fig.~\ref{fig:GlobalFits}.
These results show that the Pad\'e parametrizations provide an accurate and numerically stable description of the data, while avoiding extrapolation instabilities and unphysical behavior at large~$|t|$. The shaded bands illustrate the intrinsic sensitivity of the Pad\'e representation to controlled coefficient variations, as discussed in Sec.~\ref{sec:sec6}

From the QCD perspective, the resulting Pad\'e parametrizations exhibit two important features:
(i) a smooth behavior in the low-$t$ region, consistent with analyticity and the constraints imposed by the Regge-inspired analysis of Sec.~\ref{sec:sec5}, and
(ii) the correct large-$|t|$ power-law fall-off in accordance with perturbative QCD counting rules.
Together, these properties indicate that the Pad\'e parametrizations not only interpolate the available experimental measurements in a controlled manner, but also provide theoretically consistent analytic forms that can be reliably employed in subsequent phenomenological applications.

\section*{acknowledgements}

The work of P.M. was supported by the Ministerio de Ciencia e Innovaci\'on under grant PID2023-146142NB-I00, by the Secretaria d'Universitats i Recerca del Departament d'Empresa i Coneixement de la Generalitat de Catalunya under grant 2021 SGR 00649, and by the Spanish Ministry of Science and Innovation (MICINN) through the State Research Agency under the Severo Ochoa Centres of Excellence Programme 2025--2029 (CEX2024-001442-S). IFAE is partially funded by the CERCA program of the Generalitat de Catalunya. This work was also supported by Shahrood University of Technology. 

%
%%%%%%%%%%%%%%%%%%%%%%%%%%%%%%%%

\end{document}